\documentclass[preprint]{aastex}
\usepackage{graphics}
\usepackage{url}
\usepackage[section]{placeins}

\begin{document}

\title{Using Dimers to Measure Biosignatures and Atmospheric Pressure for Terrestrial Exoplanets}
\author{Amit Misra\altaffilmark{1,2,3}}
\affil{Box 351580, UW}
\affil{Seattle, WA 98195-1580}
\affil{Phone: 206-616-4549}
\affil{Fax: 206-685-0403}
\email{amit0@astro.washington.edu}
\and
\author{Victoria Meadows\altaffilmark{1,2,3}, Mark Claire\altaffilmark{4,2,6} \& Dave Crisp\altaffilmark{5,2}}
\altaffiltext{1}{Univ. of Washington Astronomy Dept., Seattle, Washington, USA}
\altaffiltext{2}{NAI Virtual Planetary Laboratory, Seattle, Washington, USA}
\altaffiltext{3}{Univ. of Washington Astrobiology Program, Seattle, Washington, USA}
\altaffiltext{4}{Dept. of Earth and Environmental Sciences, University of St Andrews, Fife, Scotland}
\altaffiltext{5}{Jet Propulsion Laboratory, California Institute of Technology, Pasadena, California, USA}
\altaffiltext{6}{Blue Marble Space Institute of Science, Seattle, WA, USA}
\keywords{Remote Sensing, Extrasolar terrestrial planets, Habitability, Radiative transfer}

\begin{abstract}

We present a new method to probe atmospheric pressure on Earthlike planets using (O$_2$-O$_2$) dimers in the near-infrared. We also show that dimer features could be the most readily detectable biosignatures for Earthlike atmospheres, and may even be detectable in transit transmission with the James Webb Space Telescope (JWST). The absorption by dimers changes more rapidly with pressure and density than that of monomers, and can therefore provide additional information about atmospheric pressures.  By comparing the absorption strengths of rotational and vibrational features to the absorption strengths of dimer features, we show that in some cases it may be possible to estimate the pressure at the reflecting surface of a planet. This method is demonstrated by using the O$_2$ A band and the 1.06 $\mu$m dimer feature, either in transmission or reflected spectra. It works best for planets around M dwarfs with atmospheric pressures between 0.1 and 10 bars, and for O$_2$ volume mixing ratios above 50\% of Earth's present day level. Furthermore, unlike observations of Rayleigh scattering, this method can be used at wavelengths longer than 0.6 $\mu$m, and is therefore potentially applicable, although challenging, to near-term planet characterization missions such as JWST.  We have also performed detectability studies for JWST transit transmission spectroscopy and find that the 1.06 $\mu$m and 1.27 $\mu$m dimer features could be detectable (SNR$>$3) for an Earth-analog orbiting an M5V star at a distance of 5 pc. The detection of these features could provide a constraint on the atmospheric pressure of an exoplanet, and serve as biosignatures for oxygenic photosynthesis. We have calculated the required signal-to-noise ratios to detect and characterize O$_2$ monomer and dimer features in direct imaging reflected spectra and find that signal-to-noise ratios greater than 10 at a spectral resolving power of R=100 would be required.
\end{abstract}

\section{Introduction}

Atmospheric pressure is a fundamental parameter for characterizing the environment and habitability of an extrasolar planet. Water's stability on a planetary surface as a liquid depends on both the surface temperature and pressure.  While the freezing point of water is not strongly dependent on pressure, pressure does affect water's boiling point and sublimation.  Thus, a reliable estimate of the surface pressure is an important part of the measurement suite required to determine the habitability of an exoplanet.  
 
Despite the importance of atmospheric pressure, current proposed methods for measuring pressure using remote-sensing techniques that could be applicable to exoplanet atmospheres are challenging.  The existing techniques include the use of Rayleigh scattering \citep{traub10} or the widths of individual absorption lines \citep{kaplan64, gray66} or absorption bands \citep{ign09, chamb06, spiga07, chamb13}.  The presence and location of a blue Rayleigh scattering tail in a spectrum can provide information about the existence and pressure of an atmosphere.  However, strong blue absorbers in the atmosphere (e.g. O$_3$, SO$_2$, NO$_2$ and many others) or surface features can mask this tail \citep{crow11}.  Furthermore, the Rayleigh scattering tail is most prominent shortward of 0.6 $\mu$m, below the short wavelength cutoff of the James Webb Space Telescope (JWST) \citep{gardner06}. Lastly, planets around M dwarfs are likely to be the first to be characterized \citep{deming09}, and M dwarfs have relatively less visible-flux to Rayleigh scatter than solar-type stars. The Rayleigh tail would be more difficult to detect and characterize for planets orbiting stars of this stellar class.

It is also possible to use the widths of absorption features to estimate pressure.  Pressure increases the widths of vibration rotation lines of gases. This method has been successful for the Earth using high-resolution spectra of the O$_2$ A band \citep{barton86, mitchell87, crisp12}, for Mars using CO$_2$ features near 2 $\mu$m \citep{gray66, chamb06, forget07, spiga07} and the cloud tops of Venus using the 1.6 $\mu$m CO$_2$ band \citep{ign09}. This method provides unambiguous results when the spectral resolution is sufficiently high to resolve the profiles of individual spectral lines.  It can also be used at lower spectral resolution, but requires prior knowledge of the mixing ratio of the absorbing gas. 
 
Here we explore the feasibility of a new method to directly measure the pressure of an Earth-like atmosphere that combines the absorption features of dimers with those of monomer vibration-rotation bands to yield estimates of the atmospheric pressures even when the mixing ratio of the monomer is uncertain. Previous pressure estimates using dimer absorption have been made for the cloud tops of Earth, but these techniques have required prior knowledge of the gas mixing ratio profile \citep{acaretta04}. Dimers are bound or quasi-bound states between two molecules driven together by molecular interactions. For example, the  O$_2$-O$_2$ or O$_4$ dimer consists of two O$_2$ molecules temporarily bound to each other by Van der Waals forces.  This dimer has its own rotational and vibrational modes, and produces spectral features distinct from its constituent O$_2$ monomers. Additionally, absorption from dimer molecules is more sensitive to pressure than that of monomers. The optical depth (how much absorption occurs) for dimers and monomers can be expressed by the following equations:
\begin{eqnarray}
d\tau_{monomer} &=& \sigma \rho dl  = \sigma P/T  dl\\
d\tau_{dimer} &=& k \rho^2 dl = k P^2/T^2 dl
\end{eqnarray}
where $d\tau_{monomer}$ and $d\tau_{dimer}$ are the monomer and dimer differential optical depths,  $\sigma$ is the monomer cross section, $\rho$ is the number density of the gas, $k$ is the dimer cross section, $P$ is the pressure, $T$ is the temperature, and $dl$ is the path length. While the monomer (e.g. O$_2$)  optical depth is directly proportional to pressure, the optical depth of the dimer (e.g. O$_2$-O$_2$) is dependent on the square of the density (and hence square of the pressure).  This difference in pressure dependence allows us to estimate atmospheric pressure by comparing the dimer and monomer absorption features

For an oxygen-rich, Earth-like atmosphere, the best combination of bands to use for pressure determination at near infrared wavelengths ($>$0.6 $\mu$m) would be the (O$_2$) A band at 0.76 $\mu$m, and the 1.06 $\mu$m O$_4$ dimer band.  The 0.76 $\mu$m O$_2$ A band is the strongest O$_2$ feature  in the visible-near-infrared spectral region, and is found in a relatively clean region of the spectrum between two water vapor bands.  We have chosen the dimer feature at 1.06 $\mu$m as the likely best option, due to its combination of band strength and its location in a relatively uncluttered region of the planetary spectrum.   Other O$_4$ features overlap with water features (dimer feature between 5.5 and 7 $\mu$m) or O$_2$ vibration-rotation bands (0.63 $\mu$m, 0.76 $\mu$m and 1.27 $\mu$m dimer features) or are weaker than the 1.06 $\mu$m dimer feature (0.477 $\mu$m and 0.57 $\mu$m dimer features). Nevertheless, some other features, in particular the strong 1.27 $\mu$m feature, could be used if the 1.06 $\mu$m dimer feature is not detectable.

In the proposed technique, the O$_2$ 0.76$\mu$m (monomer) band is used to provide an estimate of the atmospheric concentration of O$_2$, and combined with the O$_4$ 1.06 $\mu$m (dimer) band to constrain the atmospheric pressure.    This method can be used with either transmission spectroscopy, or directly-detected reflection or emission spectra, and serves as a complement to pressure determination techniques such as Rayleigh scattering, which only work in the visible. Although the proof of concept is shown here with oxygen, this technique is not limited to the oxygen dimer in an Earth-like atmosphere in the visible to near-infrared.  The same technique is applicable to pairs of monomer and dimer absorption features across a wider range of planetary atmospheric composition and spectral wavelength range.  
 
\section{Methods}

In this paper we generated transit transmission and direct imaging reflected spectra for cloud and aerosol-free Earthlike exoplanets. The models we have used to do this are described below.

\subsection{Transmission Spectroscopy Model}

\subsubsection{Model Overview}
When an extrasolar planet transits or occults its host star, the planetary atmosphere is backlit and some of the star's light traverses the planet's atmosphere on limb trajectories.  This transmitted light can be used to characterize the planet's atmosphere \citep{seager00, brown01, hubbard01}. This has been done for a number of Jupiter- and Neptune-sized planets (e.g. \citet{char02, vidal03, pont08}) and the super-Earth/mini-Neptune GJ1214b (e.g. \citet{bean10}) 

The transmission spectroscopy model used here is based on SMART (Spectral Mapping Atmospheric Radiative Transfer) \citep{mead96, crisp97}, which is a spectrum resolving (line-by-line), multi-stream, multiple scattering radiative transfer model. We have modified SMART to generate transit transmission spectra by combining the monochromatic absorption and scattering opacities for each atmospheric layer calculated by SMART with the limb path lengths inherent in a transit transmission event.  The transmission model included gas absorption, Rayleigh scattering, interaction-induced absorption, extinction from clouds and aerosols, refraction and limb darkening. Because multiple scattering was not included in the transit transmission component of the model, we have considered only cloud and aerosol-free atmospheres for this work.

\subsubsection{Refraction} \label{sec:model-ref}

An important characteristic of the transit transmission spectroscopy model is the inclusion of refraction. As described in \citet{garcia12}, refraction sets a fundamental limit on the range of pressures that can be probed during a transit, independent of absorption and scattering. Light refracts as it passes through an atmosphere, with a larger refraction angle as higher pressures and densities are probed. For every planet-star system there will be a maximum tangent pressure in the planet's atmosphere that can be probed, because at greater pressures the light will be refracted by too large of an angle to be able to reach a distant observer during the transit.

For each tangent height in each atmosphere, we calculated the total angle of refraction for a beam of light emitted from the host star using a modified version of the method described in \citet{auer00}. Their method was developed for calculating refraction for astronomical observations on the Earth given a tangent altitude, apparent zenith angle and an atmospheric density profile. We calculated the angle of refraction over a range of zenith angles to determine if a path exists to connect the host star to a distant observer via the planetary atmosphere. Transit transmission spectroscopy cannot probe the tangent altitudes at which no such path exists.

\begin{figure}[ht]
\centering
\includegraphics[width=8cm]{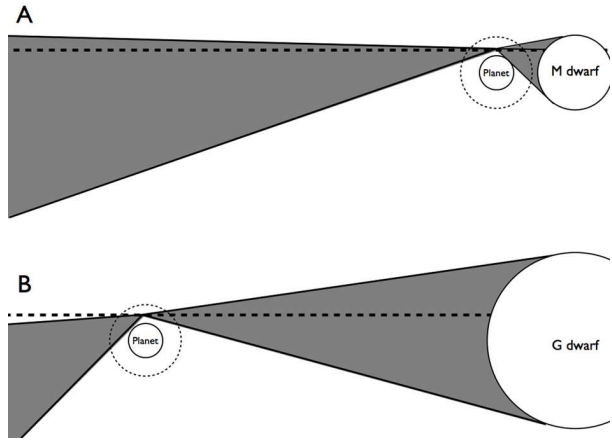}
\caption{Comparison of the effect of refraction on \textbf{(a)} an exoplanet around an M dwarf and \textbf{(b)} a planet around a Sun-like star with both planets receiving the same total flux.  The dashed circles are the planetary atmospheres. The solid lines represent different refracted paths through the atmosphere, and the dashed lines are the hypothetical paths to a distant observer observing the planet in transit transmission.  Only paths that lie exactly on that dashed line will be observed.  For the M dwarf case, there will be a path connecting the star to the observer through the atmosphere at this particular tangent height.  However, there is no path in the Sun-like star case.  This means that a transit transmission spectrum of the planet around the M dwarf can theoretically probe higher pressures of the atmosphere than in the Sun-like star case. \label{fig:refraction} }
\end{figure}

The radius of the star, planet-star distance and composition of the planet's atmosphere determine the maximum tangent pressure. The radius of the star and planet-star distance control the apparent angular size of the star from the planet's perspective. The larger the angular size, the greater the range of pressures that can be probed. For an Earth-analog orbiting a Sun-like star the angular size of the star is $\sim$0.5$^{\circ}$ while for an Earth-analog orbiting an M5 dwarf and receiving the same total flux, the angular size of the star is $\sim$2$^{\circ}$. Therefore, transit transmission spectroscopy can probe higher pressures, i.e. see deeper into the atmosphere, for the planet orbiting an M dwarf. Figure \ref{fig:refraction} shows this effect by comparing all possible paths at one tangent height for a planet around an M dwarf and a planet around a Sun-like star, with each planet receiving the same total flux. 

The composition of the atmosphere determines the refractivity (index of refraction - 1) of the atmosphere. Atmospheres with greater refractivities will have lower maximum tangent pressures. The refractivity at STP (standard temperature and pressure) can vary from $\sim$1.5 times the refractivity of air for CO$_2$ to slightly less than half the refractivity of air for H$_2$, when considering only the common bulk atmospheric gases in the solar system. Therefore, in general it will be possible to probe higher pressures for an H$_2$ atmosphere than a CO$_2$ or air atmosphere.

\subsection{Direct Imaging (Reflected) Spectroscopy Model} \label{sec:albedo}

\begin{figure}[ht]
\centering
\includegraphics[width=8cm]{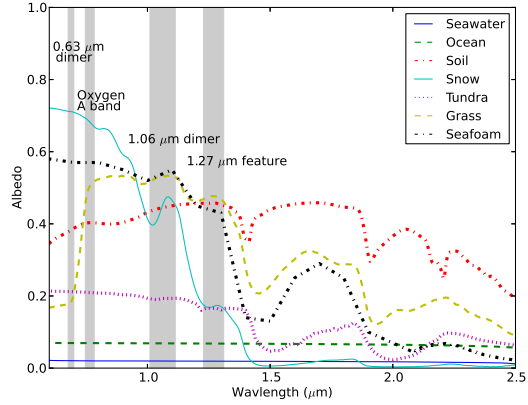}
\caption{Wavelength-dependent albedos for a variety of surfaces.  The shaded regions correspond to O$_2$ monomer and dimer bands.  For any absorption band, as long as the albedo does not vary widely within the band's wavelength range it should be possible to measure an accurate equivalent width for each feature.  The maximum variation within a band is no more than $\sim$20\%.  Therefore we do not consider surface albedo variations important for this work. \label{fig:albedo}}
\end{figure}

The reflected spectra were generated using the standard version of SMART, which can include multiple scattering from clouds and aerosols. However, to maintain consistency with the transmission spectrum model, clouds were not included in this study. The reflected spectra were generated assuming a surface with a constant albedo of 0.16, which is the average albedo of the cloud-free Earth \citep{pierrehumbert10}. We also assumed the surface is a Lambertian scatterer. Other surface types could have introduced an error into any quantitative estimates in this paper, unless explicitly included in a retrieval attempt. Figure \ref{fig:albedo} shows the wavelength-dependence of a variety of surface types from the ASTER (Advanced Spaceborne Thermal Emission and Reflection Radiometer) spectral library \citep{baldridge2009} and the USGS (United States Geological Survey) digital spectral library \citep{clark2007}. Surface albedos are  nearly constant in the bands considered here, though, for example, snow fluctuates by $\sim$20\% within the 1.06 $\mu$m dimer band. We modeled a test case with SMART with the snow surface to determine the error different surfaces can introduce. We found a difference of 15\% between the seawater and snow cases when measuring the equivalent width (defined in Section \ref{sec:measurements}) of the 1.06 $\mu$m dimer band. This difference was significantly less than the difference in measured equivalent widths for the cases considered in this paper. Therefore, we have considered any discrepancies due to variations in surface albedo to be minimal for the present work.

\subsection{Model Inputs}

\subsubsection{Model Atmospheres} \label{sec:mod-atm}

The cloud-free model atmospheres were generated by a one-dimensional (altitude) photochemical code with an extensive history in early Earth \citep{zerkle12}, modern Earth \citep{catling10} and exoplanet \citep{domagal11} research. The planetary radius and surface gravity used were the radius of the Earth (6371 km) and surface gravity of the Earth (9.87 m/s$^2$). The vertical grid consists of 200 plane-parallel layers that are each 0.5km thick in altitude, in which radiative transfer, atmospheric transport, and photochemical production and loss are solved simultaneously, subject to upper (stellar flux, atmospheric escape) and lower (volcanos and biology) boundary conditions.  The model calculates the mixing ratios of each species in each layer by solving the coupled mass-continuity/flux equations with the reverse Euler method (appropriate for stiff systems) and a variable time-stepping algorithm. Dimer concentrations were computed using estimations from quantum mechanical calculations. First, we fit the temperature dependence of the equilibrium constant for O$_2$-dimer formation using the following equations:
\begin{eqnarray}
Kp= (p(O_2)_2/(pO_2)^2)/atm^{-1} \\
Kp(T) = 2428 * T^{(-3.518)} \label{eqn:kpt}
\end{eqnarray}
where p(O$_2$)$_2$ and pO$_2$ are the dimer and monomer partial pressures and $T$ is the temperature \citep{uhlik93}. The dimer mixing ratio was then computed as:
\begin{equation}
p(O_2)_2 = Kp(T)*(pO_2)^2*P
\end{equation}
where $P$ is the pressure in atmospheres. The choice of quantum parameters in our fit of equation \ref{eqn:kpt} ensures that our calculation is a lower limit to the dimer concentrations, an assumption which matches modern atmospheric data well \citep{slanina94}.

\begin{figure}[ht]
\centering
\includegraphics[width=16.5cm]{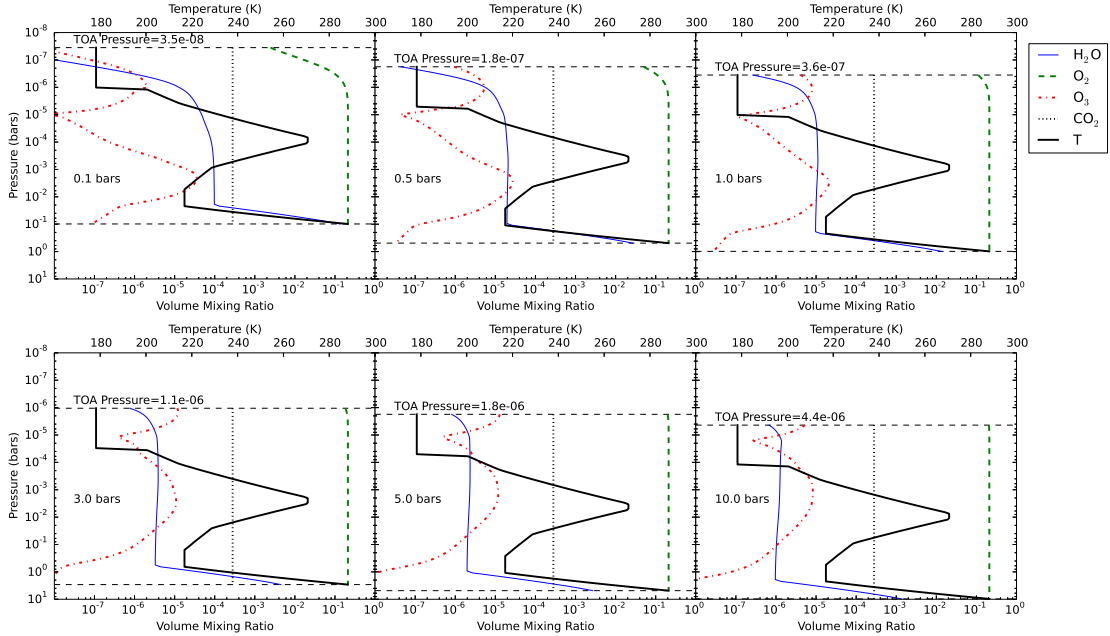}
\caption{Pressure-temperature profiles and volume mixing ratio profiles for all pressures at 1.0x PAL O$_2$. The black dashed lines represent the surface pressures and top of atmosphere pressures. We use the modern Earth temperature-altitude profile in all cases and calculate the pressures assuming hydrostatic equilibrium. The remainder of the atmosphere is N$_2$ for all cases. \label{fig:profiles}}
\end{figure}

The model atmospheres used in this study started with boundary conditions that reproduce Earth's modern atmospheric chemistry. We then replaced the solar spectrum with the M dwarf spectrum described in \ref{sec:stellar-prop} and decreased the surface albedo to 0.16 to account for cloud-free conditions. This ``modified Earth around an M dwarf'' model was then perturbed to examine changes to both total pressure and oxygen concentrations. Total atmospheric pressures of 0.1, 0.5, 1.0, 3.0, 5.0 and 10.0 bar were examined. At each of these total pressures, lower boundary conditions on O$_2$ mixing ratios were set at 0.1, 0.5, 1.0 and 2.0 times Earth’s present level. This corresponds to oxygen mixing ratios from 2\% to 42\%, which is roughly the range of oxygen concentrations experienced throughout the past ∼2.5 Gyr of Earth’s history \citep{kump08}. Boundary conditions for all other species were held fixed at modern values. Figure \ref{fig:profiles} shows the pressure-temperature profile and volume mixing ratio profiles for the 1.0 times the present atmospheric level (PAL) O$_2$ cases for spectrally active gases in the wavelength region examined here. Stable steady-state solutions were analyzed in all but the three cases with the highest O$_2$ surface partial pressures. For those three cases, we extrapolated from our converged results at lower O$_2$ mixing ratios by increasing O$_2$ concentrations, scaling the dimer concentrations with the square of the O$_2$ mixing ratio, and keeping all other gas mixing ratios constant.

For all tests presented here, we assumed Earth-like temperature and water vapor profiles, precluding the need for costly climate simulations. More specifically, we took the modern Earth temperature profile as a function of altitude, and computed the corresponding pressure levels assuming hydrostatic equilibrium. This simplification provides a useful baseline, but limits the range of validity of these results somewhat because the gas absorption cross sections and number densities depend on these atmospheric properties. The impact of this assumption is assessed in Section \ref{sec:sensitivity}. For pressure/temperature regions corresponding to Earth's mesosphere, we adopted a temperature of 180 K which likely overestimates the temperature in these regions. Dimer absorption preferentially occurs in higher pressure regions and so is insensitive to assumed conditions in the tenuous upper atmosphere. Our model grid was capped at 100km altitude ensuring optically thin conditions for nearly all species, and made allowances for CO$_2$ and N$_2$ photolysis above the upper boundary. We used the modern measured eddy-diffusion profile to simulate convective motion for all atmospheres regardless of pressure. While this simplification would affect the prediction of trace gas concentrations, all species analyzed here are either well-mixed (CO$_2$, O$_2$, N$_2$) or short-lived (dimers) so are not sensitive to changes in turbulent mixing.

\subsubsection{Absorption Line Lists and Cross Sections}

We used the HITRAN 2008 database line lists \citep{rothman09} to generate opacities in SMART. The dimer absorption cross sections for O$_2$ were taken from \citet{greenblatt90, mate99}.

\subsubsection{Stellar Properties} \label{sec:stellar-prop}

In this paper we have assumed that the planet orbits an M5V star, or an M dwarf with stellar radius of 0.20 R$_{\odot}$ and a luminosity of 0.0022 L$_{\odot}$ \citep{kalt09}.  The planet was placed at a distance of 0.047 AU, so that the total integrated incoming stellar flux was equal to the total flux the Earth receives from the Sun today. An M5 dwarf was chosen for these tests because there is a high probability that the terrestrial planets which will be the most easily characterized in the near future will be orbiting this class of star \citep{deming09}. Additionally, transit transmission spectroscopy can probe pressures up to the $\sim$1 bar level in an atmosphere for an Earth analog around an M dwarf while for the Earth around the Sun, it can only probe pressures as great as $\sim$0.2 bars.  At these pressures and lower, the 1.06 $\mu$m dimer feature is very weak, even at 200\% PAL O$_2$.  Thus, using Earth-like atmospheres around an M dwarf instead of around the Sun provides a better demonstration of the pressure-dependence of dimer features. 

To simulate the spectrum of the star, we used a Phoenix NextGen synthetic spectrum \citep{hauschildt99} for all wavelengths greater than $\sim$300 nm. For the shorter wavelengths we used the UV spectrum of AD Leo \citep{segura05}. The Phoenix spectrum was normalized so that the total integrated flux was equal to 1373 W/m$^2$, which is the integrated flux the Earth receives from the Sun. The AD Leo spectrum from \citet{segura05} was left unchanged, as it was already normalized to equal the amount of flux a planet near the inner edge of the habitable zone would receive. 

\subsubsection{Spectral Resolution}

We have used a spectral resolving power of 100 to provide relevance to the James Webb Space Telescope (JWST), which will provide new opportunities for characterizing the atmospheres of transiting exoplanets \citep{laf13}. Once the JWST is launched, the Near-Infrared Spectrograph (NIRSPEC) will provide spectra with a spectral resolving power (R=$\frac{\lambda}{\Delta \lambda}$) of $\simeq$100 between 0.6 and 5.0 $\mu$m in single prism mode \citep{kohler05}. We examine the effect of varying the spectral resolution in Sections \ref{sec:resolution1} and \ref{sec:resolution2}.

\subsection{Absorption Strength Measurements} \label{sec:measurements}

To make a quantitative estimate of absorption strengths we measured equivalent widths for the reflected spectra and measured parts per million (ppm) differences in flux for the transmission spectra.  Equivalent widths were calculated using the following equation:
\begin{equation}
W = \int (1 - F_{\lambda}/F_0) d\lambda
\end{equation}
where $W$ is the equivalent width, F$_{\lambda}$ is the flux at each wavelength $\lambda$, and F$_0$ is the continuum flux at each wavelength. To obtain the equivalent widths, we first measured the area of the spectral band below the continuum.  For each absorption band, we define the continuum by hand. For the O$_2$ A band, the continuum was assumed to be linear with wavelength, and for the 1.06 $\mu$m feature the continuum was assumed to be constant with wavelength because in several of the cases the continuum at the longer wavelengths was difficult to define due to H$_2$O absorption. The equivalent width is the width, in units of wavelength, of a rectangle measured from the continuum to the level of zero flux with the same total area as the spectral band. We could not use this type of measurement for the transmission spectra because of the difficulty in defining the zero flux level.  Therefore, we quantified the absorption strengths for the transmission spectra by measuring the change in flux from the continuum to the point of greatest absorption within a band. 

\subsection{Detectability Calculations}

We performed detectability studies \citep{kalt09, deming09, belu11, rauer11} for the model spectra, assuming the exoplanet-star system is at a distance of 5 pc. We calculated the expected signal-to-noise ratio (SNR$_{star}$) using the JWST Exposure Time Calculator (ETC) \footnote{http://jwstetc.stsci.edu}. The JWST ETC includes background noise from sky, dark, thermal and zodiacal sources along with read-out noise and photon noise. The estimates provided are expected to be within 20\% of the mission requirements. We note that for all the cases considered, the noise is dominated by photon noise. We do not include noise from detector intrapixel variations \citep{deming09}, but in principle calibration time could be devoted to mapping the pixels, as has been done with Spitzer Space Telescope Infrared Array Camera \citep{carey12}. We also note that noise levels within $\sim$20\% of the photon noise limit have been obtained for transit transmission spectra with HST in spatial scan mode, wherein the target star is trailed during each exposure by telescope motion perpendicular to the direction of dispersion \citep{deming13, wakeford13}. Spatial scan mode is being considered for JWST, and so assuming photon-limited noise in our calculations, while optimistic, provides a reasonable estimate of the detectability of absorption features (Drake Deming, private communication, October 13, 2013).

We assumed that every possible transit is observed in JWST's 5 year mission lifetime, ignoring decreases in integration time due to non-zero impact parameters (where the planet does not traverse the center of the stellar disk and therefore has a lower transit duration) and limits on visibility based on the ecliptic latitude of the exoplanet (see \citet{belu11}). For the Earth-Sun analog, this corresponds to a total integration time of $\sim$2.3*10$^5$ seconds, and for the Earth-M5V analog, an integration time of $\sim$10$^6$ seconds. We normalize the Solar spectrum (using the Phoenix G2V model available on the site) to a Johnson V magnitude of 3.32 and the M5V spectrum to 2.1*10$^{-13}$ erg cm$^{-2}$ s$^{-1}$ \AA$^{-1}$ at 1 $\mu$m.

At each wavelength within an absorption band we measured the signal as the magnitude of the difference from the continuum flux. The noise is expressed in parts per million (ppm, 10$^6$/SNR$_{star}$) at each wavelength. The final SNR of the transit transmission spectrum is the square root of the sum of the squares of the SNR at each wavelength in the absorption band divided by $\sqrt2$, which is included because the transit transmission spectrum must be calibrated against the out of transit spectrum of the star.

We also performed detectability studies for the direct imaging reflected spectra that could be relevant to proposed direct imaging planet detection and characterization missions. For these calculations, we did not use an instrument simulator because the exact specifications for these missions are not currently defined. For each absorption band, we calculated two SNRs, one for detecting the spectral feature (SNR$_D$) and one for measuring the flux at the center of the band to a precision of 3$\sigma$ (SNR$_P$). We calculated two SNRs because obtaining information about pressure from a spectral feature requires more than detection; it also requires a quantitative estimate of the strength of that spectral feature. To calculate SNR$_D$, we divided the reflected flux by the stellar flux, defined a continuum and calculated the signal as the difference between the continuum and the normalized reflected flux. We assumed the noise is constant over the entire absorption band, and then calculated the noise level required to detect the spectral feature with a SNR$_{band}$ of 3 in the absorption band. The final SNR$_D$ is the mean of the continuum reflected flux level divided by the calculated noise. To calculate SNR$_P$, we selected the wavelength within the band with the lowest radiance. We set the value of a second noise level as the lowest radiance divided by 3. SNR$_P$ is the continuum flux level divided by the noise required to obtain a SNR of 3 at the lowest radiance in the absorption band.

\section{Results} \label{sec:results}

\subsection{Transit Transmission Spectra}

\begin{figure}[ht]
\centering
\includegraphics[width=8cm]{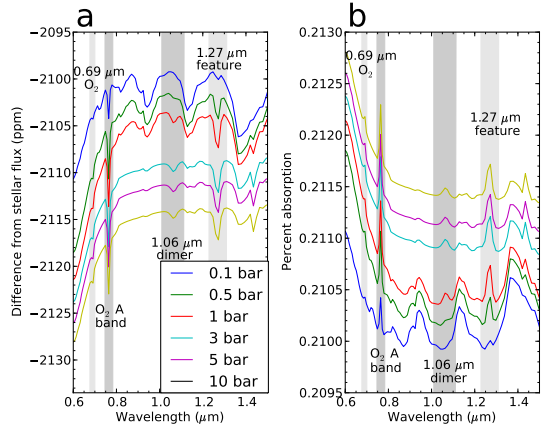}
\caption{Transmission spectra of an Earth-like atmosphere with different total atmospheric pressures.  \textbf{(a)} Difference in flux from the stellar flux.  \textbf{(b)} Percent of stellar flux absorbed by the atmosphere.  The 1.06 $\mu$m dimer feature is strong only in the spectra corresponding to atmospheric pressures greater than $\sim$0.5 bar.  The spectra for atmospheres with pressure $\geq$1 bar are nearly identical except for an offset because there is a fundamental limit on which heights in an atmosphere can be probed using transmission spectroscopy.  For an Earth-like atmosphere around the M5V used here, only the top 0.9 bars can be probed for all atmospheres at all wavelengths.  This pressure corresponds to an altitude of $\sim$1 km above the surface for a 1 bar atmosphere, and 19 km above the surface for a 10 bar atmosphere. The offset is due to the flux blocked by layers in the atmosphere with pressure greater than 0.9 bars.\label{fig:plottran}}
\end{figure}

We have generated transit transmission spectra and direct imaging spectra for atmospheres with O$_2$ concentrations of 10\%, 50\%, 100\% and 200\% of the PAL and pressures of 0.1, 0.5, 1, 3, 5 and 10 bars. Figure \ref{fig:plottran} shows the resulting transit transmission spectra for 100\% PAL O$_2$. Transmission spectra of atmospheres with pressures $\geq$1 bar are nearly identical for a given O$_2$ concentration because refraction limits to depth of penetration to $<$1 bar. The dimer feature is weak for pressures of 0.1 and 0.5 bars.

\begin{figure}[ht]
\centering
\includegraphics[width=8cm]{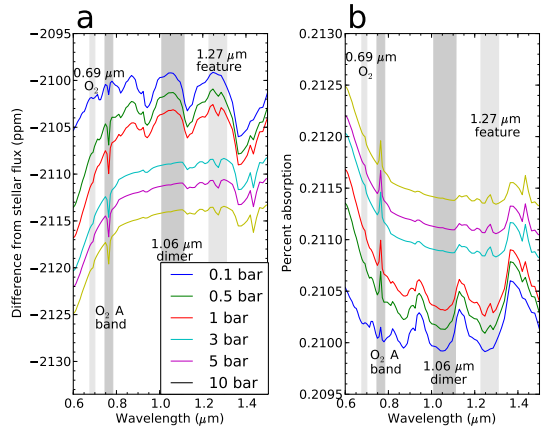}
\caption{Same as Figure \ref{fig:plottran} but for 10\% of the present day level of O$_2$.  The dimer features do not appear at all because the O$_2$ concentration is too low. \label{fig:plottran-0.1}}
\end{figure}

\begin{figure}[ht]
\centering
\includegraphics[width=8cm]{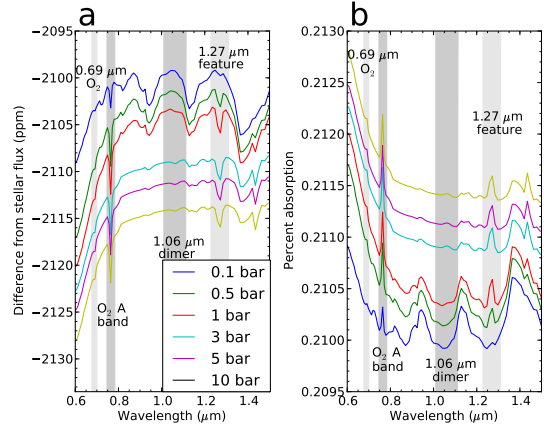}
\caption{Same as Figure \ref{fig:plottran} but for 50\% of the present day level of O$_2$. The 1.06 $\mu$m dimer feature is very weak, and is still weak for the highest surface pressure atmospheres.  \label{fig:plottran-0.5}}
\end{figure}

\begin{figure}[ht]
\centering
\includegraphics[width=8cm]{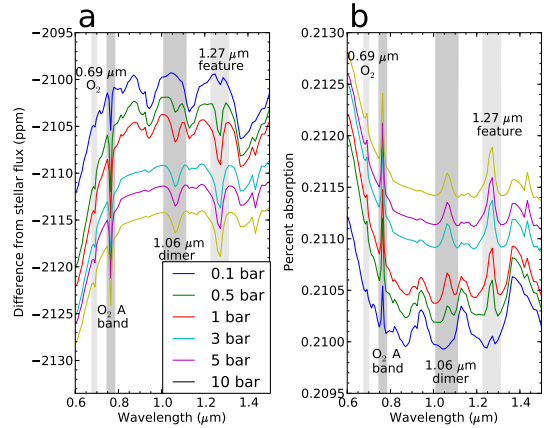}
\caption{Same as Figure \ref{fig:plottran} but for 200\% of the present day level of O$_2$. The 1.06 $\mu$m dimer feature is very strong in every case except for the 0.1 bar atmosphere.\label{fig:plottran-2.0}}
\end{figure}

Figures \ref{fig:plottran-0.1}, \ref{fig:plottran-0.5}, and \ref{fig:plottran-2.0} show the resulting transit transmission spectra for O$_2$ concentrations of 10\%, 50\% and 200\% PAL. The 1.06 $\mu$m dimer feature is very weak at all pressures for O$_2$ concentrations at 10\% PAL and weak for 50\% PAL O$_2$, while it is very strong for pressures above 0.1 bars at 200\%  PAL O$_2$.

\subsection{Reflected Spectra}

\begin{figure}[ht]
\centering
\includegraphics[width=8cm]{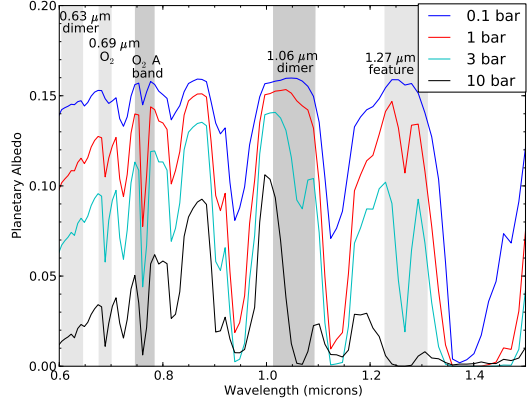}
\caption{Reflected spectra for Earth-like atmospheres with 100\% PAL O$_2$ but with different total atmospheric pressures. In the reflected spectrum there is no fundamental limit to which pressures can be probed in an atmosphere.  Assuming a cloud-free case, it is possible to probe the surface layers of the atmosphere.  The 1.06 $\mu$m dimer feature is fairly weak in the present day Earth's atmosphere, but it is a very strong feature in atmospheres with greater pressures. The 1.27 $\mu$m feature is very strong in most of the spectra.  \label{fig:plotrad}}
\end{figure}

Figure \ref{fig:plotrad} shows the reflected spectra for modern O$_2$ mixing ratios in atmospheres with a range of total pressure from 0.1 to 10 bars. The 0.5 and 5.0 bar cases are omitted from the plots to increase clarity, but are still included in the equivalent width and SNR calculations. The 1.06 $\mu$m dimer feature is extremely weak for the present-day atmosphere, but in contrast to its behavior in the transmission spectra, it is a very prominent feature for the 3, 5 and 10 bar atmospheres.

\begin{figure}[ht]
\centering
\includegraphics[width=8cm]{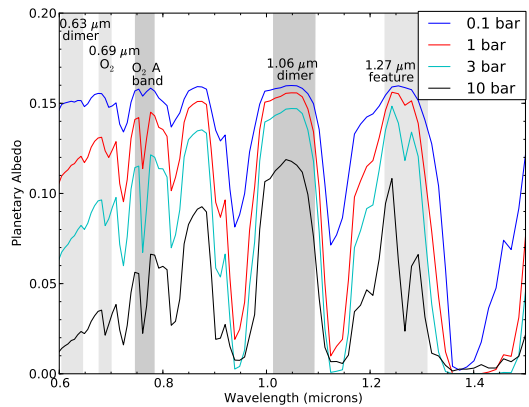}
\caption{Same as Figure \ref{fig:plotrad} but for an O$_2$ concentration of 10\% PAL.  For this amount of O$_2$, the 1.06 $\mu$m dimer feature is very weak. However, the 1.27 $\mu$m feature is still quite strong.\label{fig:plotrad01}}
\end{figure}

\begin{figure}[ht]
\centering
\includegraphics[width=8cm]{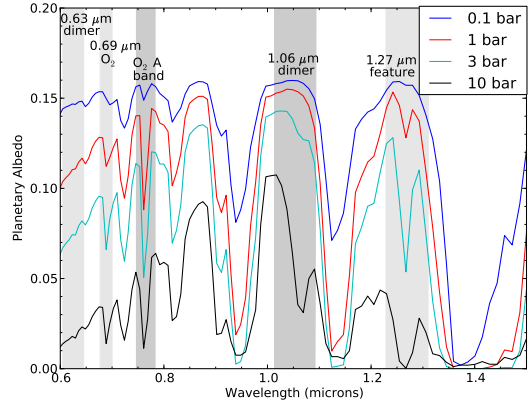}
\caption{Same as Figure \ref{fig:plotrad} but for an O$_2$ concentration of 50\% PAL.  The 1.06 $\mu$m dimer feature is strong at pressures $\ge$3 bars. \label{fig:plotrad05}}
\end{figure}

\begin{figure}[ht]
\centering
\includegraphics[width=8cm]{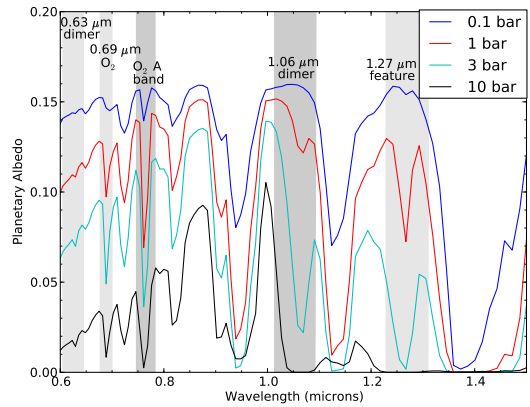}
\caption{Same as Figure \ref{fig:plotrad} but for an O$_2$ concentration of 200\% PAL.  The dimer features are much stronger with more O$_2$ in the atmosphere, as expected.\label{fig:plotrad2}}
\end{figure}

Figures \ref{fig:plotrad01}, \ref{fig:plotrad05} and \ref{fig:plotrad2} show the reflected spectra at pressures between 0.1 and 10 bars for 10\%, 50\%, and 200\% PAL of O$_2$, respectively.  In the 10\% PAL case, the O$_2$ dimer feature at 1.06 $\mu$m is very weak because the total amount of O$_2$ is very low, even for a 10 bar atmosphere.  The 1.06 $\mu$m dimer feature is stronger in the 3, 5 and 10 bar cases for atmospheres with 50\% PAL O$_2$.  Finally for the 200\% O$_2$ atmospheres, the 1.06 $\mu$m dimer feature is one of the strongest spectral features, even in the 1 bar atmosphere.

\subsection{Quantitative Absorption Measurements}

\begin{figure}[ht]
\centering
\includegraphics[width=8cm]{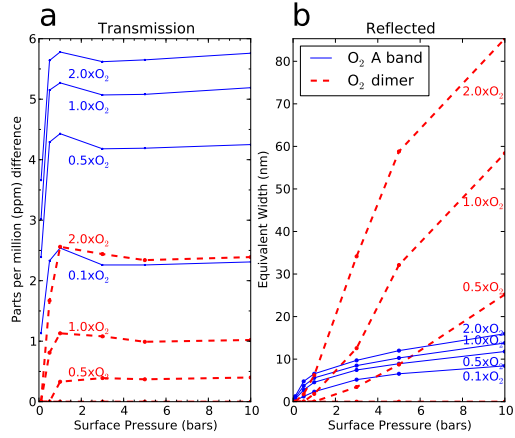}
\caption{\textbf{(a)}Flux differences (in ppm) for transit transmission spectra and \textbf{(b)}equivalent widths (in nm) for reflected spectra for the O$_2$ A band and the 1.06 $\mu$m dimer feature at various pressures and O$_2$ concentrations. In transmission, for the Earth-like planet orbiting an M5V star considered here, only the top 0.9 bars can be probed, meaning the dimer and A band equivalent widths are roughly constant with pressure above 1.0 bars for a given O$_2$ concentration.  In comparison to the reflected spectra, the dimer equivalent widths are extremely sensitive to pressure. \label{fig:equivwidths}}
\end{figure}

Figure \ref{fig:equivwidths} shows the flux change for the transmission spectra and equivalent widths for reflected spectra at different pressures and O$_2$ concentrations for the O$_2$ A band and the 1.06 $\mu$m dimer feature.  For a given O$_2$ concentration the transit transmission flux differences for the O$_2$ A band are roughly constant for pressures $\geq$1 bar due to refraction. The 1.06 $\mu$m dimer feature flux differences increase slightly with pressure but are also constant for pressures $\geq$1 bar for a given O$_2$ concentration. The dimer feature does not appear in transmission for cases with 10\% PAL O$_2$.

For the direct imaging (reflected) spectra, both the O$_2$ A band and 1.06 $\mu$m dimer feature equivalent widths increase with pressure and increased O$_2$ concentrations. However, the dimer feature equivalent widths are much more sensitive to pressure. At higher pressures the dimer feature is strong except for cases with 10\% PAL O$_2$, in which the 1.06 $\mu$m dimer feature is too weak to quantify. 

\begin{figure}[ht]
\centering
\includegraphics[width=16.5cm]{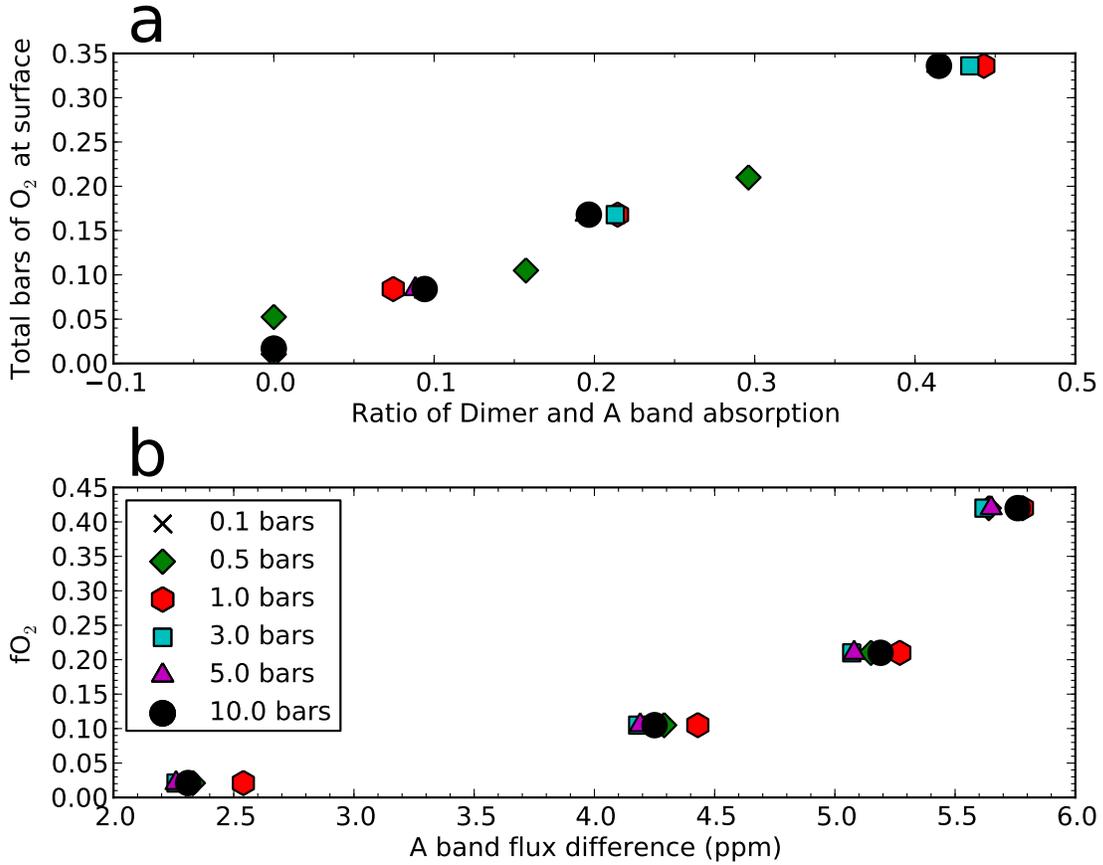}
\caption{This plot shows how one could determine a lower limit on the surface or cloud-top pressure using only a transmission spectrum.\textbf{(a)} O$_2$ partial pressure at maximum tangent pressure vs. the ratio of the dimer feature and O$_2$ A band flux difference ratio. The ratio can be used to determine the O$_2$ partial pressure. \textbf{(b)} fO$_2$ (or O$_2$ mixing ratio) vs A band flux difference. The A band flux difference is roughly constant for a given O$_2$ mixing ratio. \label{fig:pressure-tran-only}}
\end{figure}

\begin{figure}[ht]
\centering
\includegraphics[width=8cm]{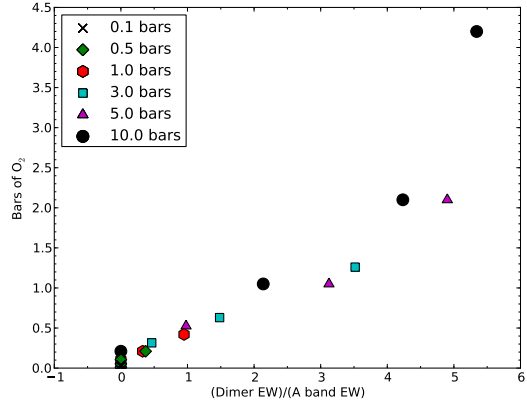}
\caption{Plot of total atmospheric O$_2$ vs the ratio of the 1.06 $\mu$m dimer feature equivalent width to the O$_2$ A band equivalent width for the reflected spectrum. There is a trend between this ratio and the total amount of O$_2$.  This could be used as a way to estimate the pressure if a transmission spectrum is not available. \label{fig:refl-only-toto2}}
\end{figure}

\begin{figure}
 \centering
\includegraphics[width=8cm]{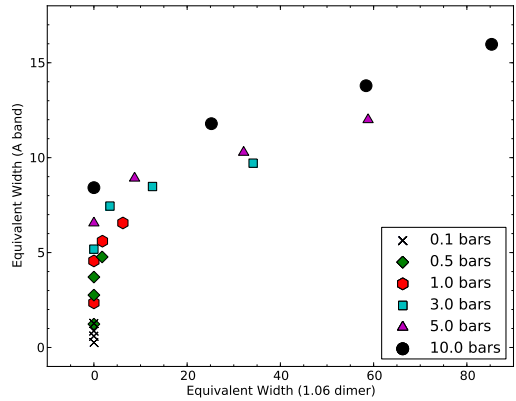}
\caption{Atmospheric pressure as a function of the dimer and A band equivalent widths in the reflected spectrum. It will be possible to set a lower bound on the pressure with only the O$_2$ A band equivalent width and the 1.06 $\mu$m dimer equivalent width. \label{fig:refl-only} }
\end{figure}

Figures \ref{fig:pressure-tran-only}-\ref{fig:refl-only} show the relationships between the quantitative absorption measurements described above and atmospheric quantities including the O$_2$ mixing ratio and the O$_2$ partial pressure at the surface. The ppm flux difference measured in transit transmission for the O$_2$ A band could be used to constrain the O$_2$ mixing ratio, as shown in Figure \ref{fig:pressure-tran-only}B. The O$_2$ partial pressure at the surface can be estimated using  the ratio between the 1.06 $\mu$m dimer and O$_2$ A band absorption measurements. These ratios are shown in Figure \ref{fig:pressure-tran-only}a using ppm flux differences, and in Figure \ref{fig:refl-only-toto2} using equivalent widths. The O$_2$ mixing ratio and O$_2$ partial pressure at the surface can be combined to provide a unique estimate of the surface pressure of the planet. A more detailed description of the pressure measurement technique is given in Section \ref{sec:measure}.

\subsection{Sensitivity Tests} \label{sec:sensitivity}

To quantify the errors introduced by assuming the modern day temperature profile, we generated spectra to test the sensitivity of our models to changes in the temperature profile and in changes to the water vapor profiles. We compared our 1.0 bar, 1.0x PAL O$_2$ spectra to a spectrum generated using the same volume mixing ratio profiles but with an isothermal atmosphere at 250 K. We also compared our 1.0 bar, 1.0x PAL O$_2$ transit transmission spectrum to spectra generated with atmospheres with 0.1 and 10.0 times the H$_2$O levels. We perform a similar comparison for the 1.0 bar, 2.0x PAL O$_2$ reflected spectrum.

\begin{figure}[ht]
\centering
\includegraphics[width=8cm]{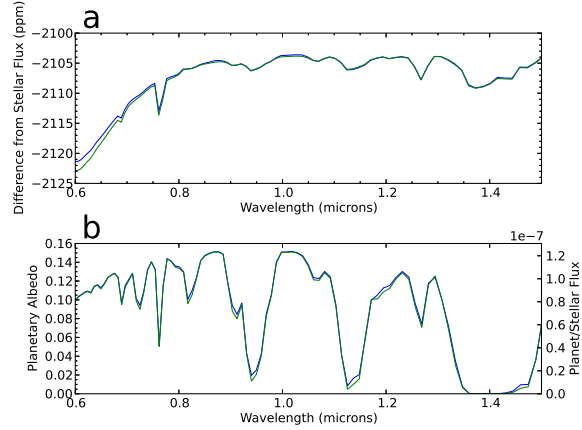}
\caption{(\textbf{a}) Transit transmission spectra for the Earth (blue) and an isothermal atmosphere at 250K with the same pressure-composition profiles (green).  (\textbf{b}) Reflected spectra for an Earth analog with 2.0x PAL O$_2$ (blue) and an isothermal atmosphere at 250K with the same pressure-composition profiles (green). In both cases, the spectra for the calculated temperature-pressure profile and the isothermal profile are very similar, showing that the spectra are not strongly dependent on the temperature profile. }
\label{fig:iso-comp}
\end{figure}

Figure \ref{fig:iso-comp} shows the sensitivity of the spectra to the temperature profile, with our Earth-like profile and an isothermal approximation profile compared. Both the transit transmission and reflected spectra show little sensitivity to the temperature profile. The 1.06 $\mu$m dimer band also shows little sensitivity to the temperature profile, despite the dependence of the dimer optical depth on the square of the temperature because the isothermal temperature profile approximates the average temeprature of the troposphere, where the majority of dimer absorption occurs.

\begin{figure}[ht]
\centering
\includegraphics[width=8cm]{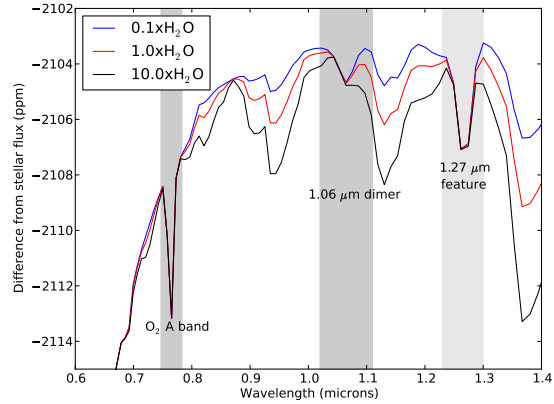}
\caption{Transit transmission spectra of a 1.0 bar, 1.0x PAL O$_2$ Earth analog with H$_2$O concentrations varying from 0.1 to 10.0x PAL H$_2$O.  While H$_2$O absorbs near the wings of the O$_2$ A band the ppm flux difference and the SNR do not change greatly as the H$_2$O concentration changes. The continuum near the 1.06 $\mu$m dimer feature is strongly affected by the increases in H$_2$O. However, the magnitude of the change in the ppm flux difference over the entire band and the SNR is less than 20\% when comparing the 1.0x PAL H$_2$O case to either the 0.1 or 10.0x H$_2$O cases. \label{fig:h2o-test-1}}
\end{figure}

\begin{figure}[ht]
\centering
\includegraphics[width=8cm]{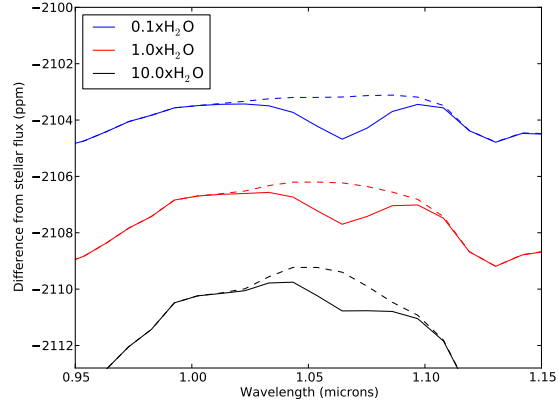}
\caption{The 1.06 $\mu$m dimer feature in transit transmission with different amounts of H$_2$O. The spectra have been artificially offset for ease of viewing. The black dashed lines are spectra that do not include any O$_2$ dimer absorption and are included to help show the continuum flux level. The total ppm flux difference and SNR for the dimer band vary by less than 20\% with respect to the 1.0xH$_2$O case.\label{fig:h2o-test-2} }
\end{figure}

\begin{figure}[ht]
\centering
\includegraphics[width=8cm]{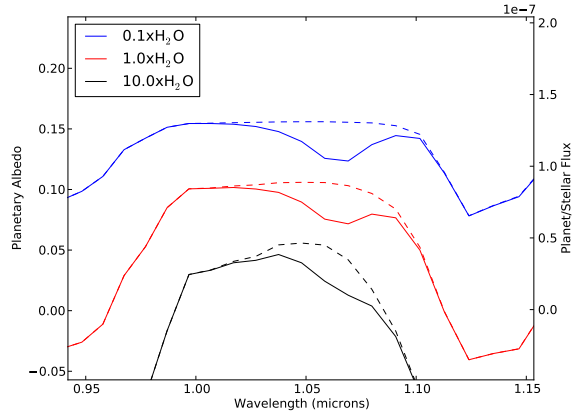}
\caption{The 1.06 $\mu$m dimer feature in the direct beam reflected spectra with varying amounts of H$_2$O for an Earth analog with a 1.0 bar, 2.0x PAL O$_2$ atmosphere.  There is an artificial offset for ease of viewing, and the black dashed lines are spectra that do not include any O$_2$ dimer absorption. As in transit transmission, the SNR of the dimer feature varies by less than 20\% with respect to the 1.0x H$_2$O case. Additionally, the SNRs for higher pressure cases should be less affected by changes in H$_2$O concentrations because of greater absorption within the dimer band.\label{fig:h2o-test-3}}
\end{figure}

Figures \ref{fig:h2o-test-1}, \ref{fig:h2o-test-2} and \ref{fig:h2o-test-3} show the results for the H$_2$O sensitivity tests. The O$_2$ A band equivalent widths and ppm flux differences are not strongly affected by changes in the H$_2$O profiles. However, changes in the H$_2$O mixing ratios affect the continuum flux in the wings of the 1.06 micron dimer feature, complicating measurements of the equivalent width of this feature. For the transit transmission spectra, the change in the total ppm flux difference (across the entire band) is less than 20\% between the 0.1 and 10.0x H$_2$O cases. For the reflected spectra, the change in the equivalent width of the 1.06 $\mu$m dimer feature is less than 20\%. For the reflected spectra, the equivalent widths can be much greater than that for the 1.0 bar, 2.0x O$_2$ case. For these greater equivalent widths, the effect of increasing or decreasing H$_2$O levels will diminish as the difference from the continuum flux (affected by H$_2$O) and the flux within the absorption band will increase.

\subsection{Detectability of Spectral Features}

Table \ref{tab:detect} shows the SNRs for observations by JWST for the O$_2$ A band, 1.06 $\mu$m feature and the 1.27 $\mu$m feature for the range of pressures and O$_2$ concentrations considered here for an Earth analog at a distance of 5 pc. The SNR are calculated assuming that every transit of an Earth analog orbiting an M5V star is observed over JWST's 5 year mission lifetime. In transit transmission, the O$_2$ A band SNRs are no greater than 1.1. The 1.06$\mu$m dimer feature is detectable at a SNR of $>$3 for many of the 2.0x PAL O$_2$ cases. The 1.27 $\mu$m feature is the most detectable O$_2$ feature in this wavelength range, with SNRs greater than 5 for many of the 1.0x PAL O$_2$ cases and greater than 7 for many of the 2.0x PAL O$_2$ cases. JWST will not be able to detect O$_2$ species for Earth-like exoplanets in secondary eclipse in the visible and near-infrared, as shown by the secondary eclipse SNR levels. Even for the highest pressure cases, the SNRs are no greater than $\sim$0.2.

Table \ref{tab:TPF} shows the SNRs necessary to detect and characterize the O$_2$ A band, 1.06 $\mu$m band and the 1.27 $\mu$m feature for the range of pressures and O$_2$ concentrations considered here for the direct imaging reflected spectra at R=100. The SNRs in Table \ref{tab:TPF} would be relevant to a direct imaging characterization mission. While these SNRs were calculated for an Earth analog orbiting an M5V star, the results should be independent of stellar spectral type because we divided out the stellar flux in our calculations. The O$_2$ A band, 1.06 $\mu$m dimer feature and 1.27 $\mu$m feature are detectable at an average SNR$_D$ of 14, 9 and 14, respectively, for the cases when the features are strong enough to identify in the model spectra. The average required SNR$_P$ to use the features for pressure estimation are 11, 31 and 34.

\subsection{Effect of Spectral Resolving Power} \label{sec:resolution1}

\begin{figure}
\centering
\includegraphics[width=8cm]{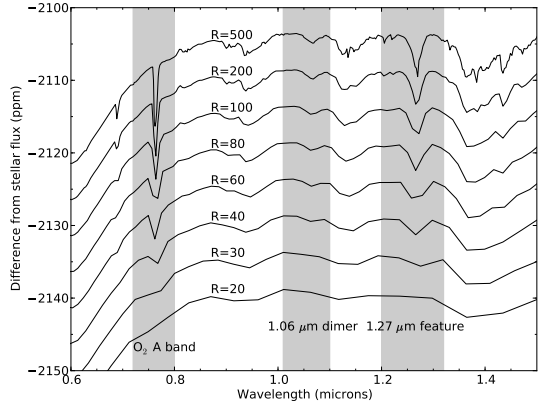}
\caption{Transit transmission spectra for the 1.0 bar, 1.0x PAL O$_2$ case at various spectral resolving powers. The wavelengths for the O$_2$ A band, the 1.06 $\mu$m dimer band, and the 1.27 $\mu$m band are highlighted. An artificial offset has been added to the spectra for ease of viewing. At the lowest resolving powers, it is difficult to identify spectral features. \label{fig:specres1}}
\end{figure}

We examined the effect of spectral resolving power on the detectability of spectral features by generating transit transmission spectra for the 1.0 bar, 1.0x PAL O$_2$ case with spectral resolving powers of 500, 200, 100, 80, 60, 40, 30 and 20. Figure \ref{fig:specres1} shows the spectra for each of these cases. We also measured the SNR of each feature at each resolving power. The SNRs were calculated assuming a noise profile equivalent to the JWST NIRSPEC noise profile, but with the noise level at each wavelength divided by (100/R)$^2$, so that the noise at each wavelength decreased as the resolving power decreased. Figure \ref{fig:specres2} shows how the total SNR in each absorption band changes with resolving power. In general, the SNRs drop off rapidly as resolving power decreases for R$<$60.

\begin{figure}
\centering
\includegraphics[width=8cm]{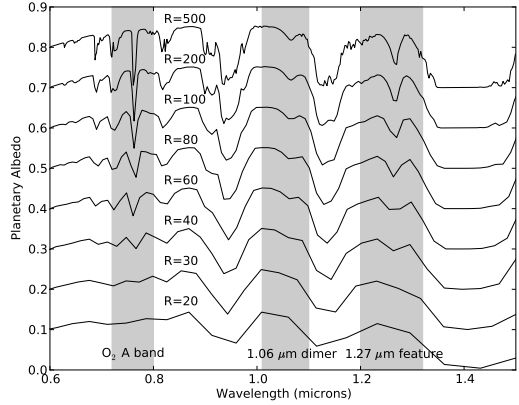}
\caption{Direct imaging reflected spectra generated at spectral resolving powers from 500 to 20. The y-axis is the planetary albedo with an arbitrary offset added for ease of viewing.  The O$_2$ A band, 1.06 $\mu$m dimer feature and 1.27 $\mu$m feature are highlighted. At resolving powers of R=30 and R=20 it is difficult to identify any O$_2$ absorption features.\label{fig:specres-imaging} }
\end{figure}

\begin{figure}
\centering
\includegraphics[width=8cm]{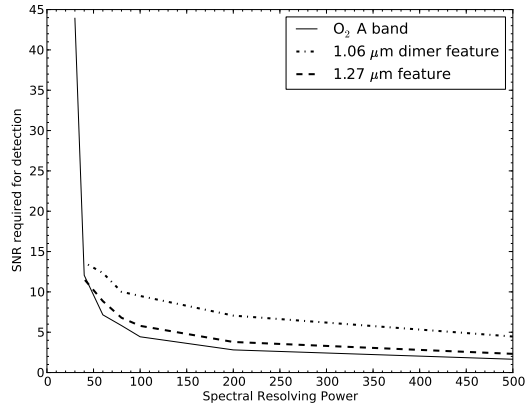}
\caption{SNRs needed to detect the O$_2$ A band, 1.06 $\mu$m dimer feature and the 1.27 $\mu$m feature for direct imaging reflected spectra. As resolving power decreases, the required SNR to detect each feature increases because the continuum level decreases, resulting in a lower overall signal in the absorption band. Furthermore, at the lowest resolutions the absorption bands cannot be resolved. Thus, this decreases the measurable signal even more, leading to an increase in the required SNR. \label{fig:specres-imaging-snrd} }
\end{figure}

\begin{figure}
\centering
\includegraphics[width=8cm]{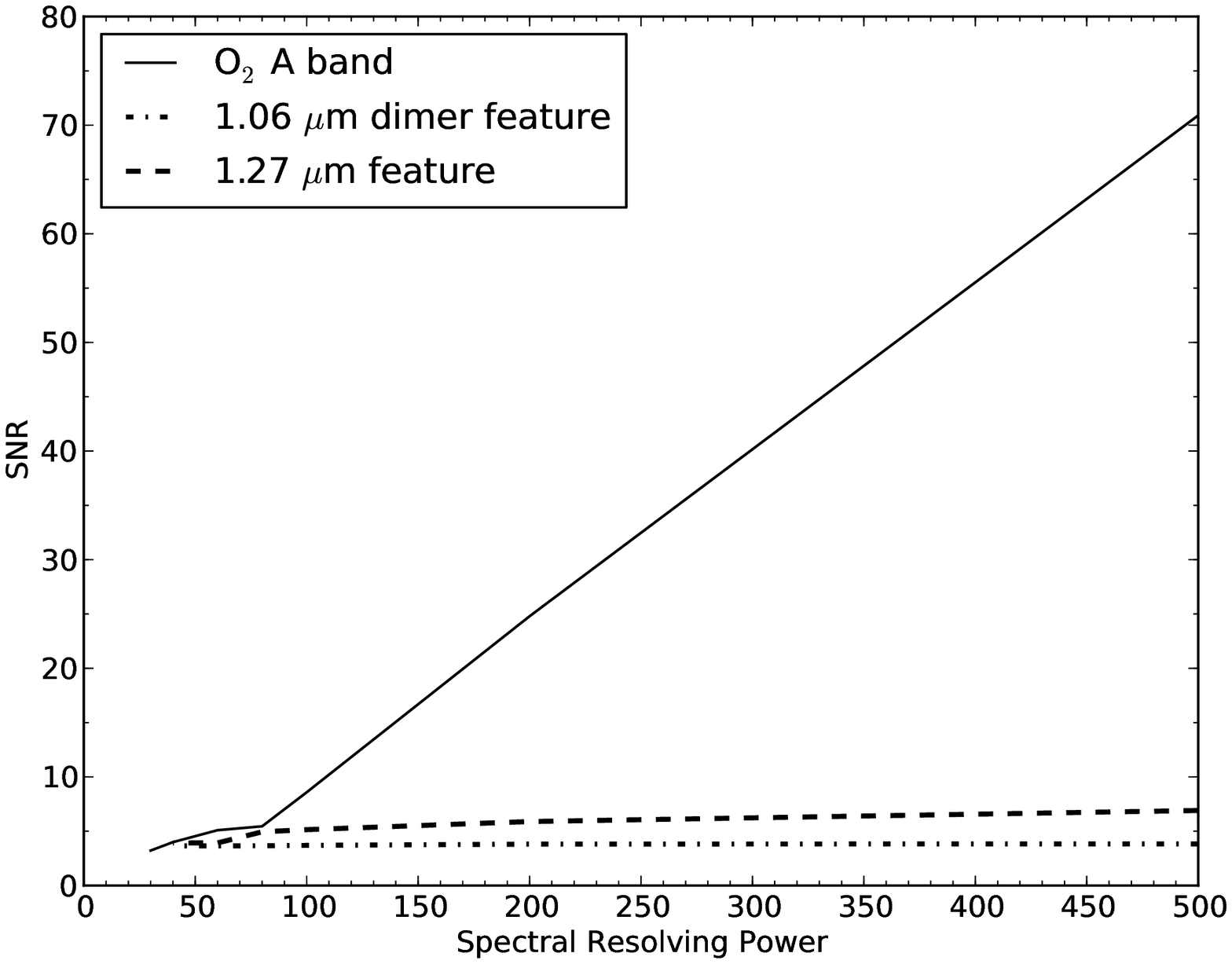}
\caption{SNRs required to quantify the flux within 3$\sigma$ in the center of the absorption band (defined as the lowest flux level) for the O$_2$ A band, 1.06 $\mu$m dimer feature and the 1.27 $\mu$m feature. The SNR required for this level of precision decreases as resolving power decreases because the flux is averaged over fewer wavelength bins, leading to an increase in the lowest flux. This decrease in required resolving power occurs for all three absorption features, though the trend is most apparent for the O$_2$ A band, as it has the narrowest spectral shape of the three features considered here. \label{fig:specres-imaging-snrp} }
\end{figure}

We also generated direct imaging reflected spectra for the 1.0 bar, 2.0x PAL O$_2$ case at resolving powers of 500, 200, 100, 80, 60, 40, 30 and 20, as shown in Figure \ref{fig:specres-imaging}. Figures \ref{fig:specres-imaging-snrd} and \ref{fig:specres-imaging-snrp} show how the SNR for detection and precision vary with resolving power for the O$_2$ A band, 1.06 $\mu$m dimer feature and the 1.27 $\mu$m feature. The SNR required to detect each feature increases as resolving power decreases. At R=20, the SNRs are not shown because no spectral features could be identified, and at R=30 only the O$_2$ A band was identified. The SNR required to quantify the flux at the center of each spectral feature decreases as resolving power decreases, because the lowest radiance level increases as the spectral resolving power decreases. 

\section{Discussion}

The 1.06 $\mu$m dimer feature is prominent in transit transmission for atmospheres with $\geq$50\% PAL O$_2$ and surface pressures $\geq$0.5 bars. It is a prominent feature in the reflected spectra for atmospheres with $\geq$50\% PAL O$_2$ and surface pressure $\geq$3 bars. For the reflected spectra, the dimer feature equivalent width is highly dependent on surface pressure when compared to the O$_2$ A band and therefore the dimer feature can be used to constrain pressure. Here we discuss how to do this for the cases investigated here.

\subsection{Pressure Measurement Technique} \label{sec:measure}

Figure \ref{fig:equivwidths} confirms that dimer absorption features are more strongly dependent on pressure than monomer features. Therefore dimer features can be combined with monomer features to determine pressure, even if the mixing ratio of the absorbing gas is not known.  With only transit transmission spectroscopy, it is impossible to probe pressures over $\sim$1 bar for the cases examined here, but it may be possible to constrain the O$_2$ mixing ratio and set a lower bound for pressure. In reflected spectra, it is possible to determine the surface partial pressure of O$_2$ and set a lower bound for pressure using the 1.06 $\mu$m dimer feature as an on/off pressure gauge. With both a transit transmission spectrum and a reflected spectrum, it should be possible to determine total atmospheric surface pressure for an Earth-like exoplanet.

\subsubsection{Transmission Spectra Pressure Measurement}

Transmission spectroscopy provides only a lower bound on the atmospheric pressure because refraction provides a fundamental limit to which pressures can be probed using this technique.  For the spectra presented here, a lower limit of $\sim$1 bar can be set for the high pressure atmospheres.  For a given O$_2$ concentration a unique estimate of the pressure can be retrieved from the ratio between the ppm flux differences of the 1.06 $\mu$m dimer feature and the O$_2$ A band.  Figure \ref{fig:pressure-tran-only}a shows the relationship between this ratio and the total amount of O$_2$ above 0.9 bars, which is the highest pressure that can be probed in this particular case.  There is a clear  trend between this ratio and the amount of O$_2$ in the atmosphere.  When combined with an O$_2$ mixing ratio this relationship can provide a quantitative estimate of a lower level of the surface or cloud-top pressure.

The O$_2$ mixing ratio can be estimated from the flux difference of the O$_2$ A band in transmission.  Figure \ref{fig:pressure-tran-only}B shows the relationship between the O$_2$ mixing ratio (which is constant throughout the atmosphere) and the O$_2$ A band flux difference. For pressure $\ge$0.1 bars the O$_2$ flux difference is roughly constant for a given O$_2$ concentration, meaning that a measurement of the O$_2$ A band flux difference should correlate with the O$_2$ mixing ratio.

\subsubsection{Reflected Spectra Pressure Measurement}

Reflected spectra alone can provide an estimate of the surface partial pressure of O$_2$ by examination of the ratio of the 1.06 $\mu$m dimer equivalent width and the O$_2$ A band equivalent width.  Figure \ref{fig:refl-only-toto2} shows the relationship between this ratio and the surface partial pressure of O$_2$. The strength of the 1.06 $\mu$m dimer feature could also be used as an on/off gauge to set a lower bound for pressure. Determining total atmospheric pressure with only a reflected spectrum is difficult due to degeneracies between the O$_2$ concentration and total atmospheric pressure, as shown in Figure \ref{fig:refl-only}. For large equivalent widths of the dimer feature, the pressure will certainly be above 1 bar. However, it is difficult to differentiate between atmospheres with the same O$_2$ surface partial pressure. Nevertheless, it appears that it is possible to set a lower limit on pressure by measuring the 1.06 $\mu$m dimer feature equivalent width. For example, a 1.06 $\mu$m dimer feature equivalent width greater than $\sim$10 nm would imply a surface pressure $>$1 bar.

\subsubsection{Pressure Measurement with both transit transmission and reflected spectra}

If both a transit transmission spectrum and a reflected spectrum are available, it should be possible to directly measure the total atmospheric surface pressure in the absence of clouds. Transit transmission spectroscopy can provide an estimate of the O$_2$ mixing ratio as described previously. A reflected spectrum can theoretically probe to the reflecting surface and therefore can be used to constrain the O$_2$ partial pressure at the surface. By combining the O$_2$ mixing ratio and O$_2$ partial pressure we can determine the total pressure at the reflecting surface, which could either be a reflective cloud layer or the physical surface of the planet.

\subsection{Relevance to Planet Characterization} \label{sec:char}

The methods described here could be used in the near future by the JWST NIRSPEC instrument, which will potentially be able to characterize transiting planets between 0.6 and 5.0 $\mu$m. The O$_2$ A band will likely not be detectable for a nearby Earth-analog with JWST. Although this feature is strong in the spectrum, the sensitivity of NIRSPEC is poor at shorter wavelengths. The 1.06 $\mu$m dimer feature is detectable at the 3$\sigma$ level in transit transmission for cases with 2.0x PAL O$_2$ and high surface pressures. Thus, the detection of the 1.06 $\mu$m dimer feature would imply a surface or cloud-top pressure greater than or equal to 1.0 bars. For cases in which the 1.06 $\mu$m dimer feature is not detectable, the 1.27 $\mu$m feature could be used to constrain the pressure. This feature is not as strongly dependent on pressure as the 1.06 $\mu$m dimer feature, but it is more detectable in all cases explored here.

TPF, Darwin or a similar direct imaging mission will be required to characterize the reflected spectra of nearby Earth analogs in the visible and near-infrared. The SNR values for secondary eclipse using JWST are all less than 1, so JWST will not be able to characterize the reflected spectra of Earth-analogs in secondary eclipse. Table \ref{tab:TPF} shows the necessary SNRs to detect and characterize spectral features for a direct imaging planet characterization mission. While the SNRs were calculated for an Earth analog orbiting an M5V star, the results should be largely independent of spectral type because we have divided the reflected flux by the stellar flux in our calculations. The required SNR values suggest that a SNR of $>$10 would be necessary to detect and quantify the O$_2$ A band, 1.06 $\mu$m dimer feature and 1.27 $\mu$m feature for a true Earth analog. However, because continuum brightness changes with pressure, a different SNR criteria would be necessary for higher pressure atmospheres. For example, the continuum brightness near the O$_2$ A band is three times lower for the 10.0 bar cases than it is for the 0.1 bar cases. For most cases, a SNR of $>$7 would likely be sufficient to set a lower limit on the surface pressure using the 1.06 $\mu$m dimer feature.

Clouds and aerosols will also affect the detectability of absorption features.  In transit transmission, clouds can effectively mask the highest pressures of the atmospheres at which dimer absorption is most prominent. However, in partially cloudy atmospheres some of the paths will probe pressures as high as the maximum tangent pressure, and thus the planetary transmission spectrum could show evidence of dimer absorption. Furthermore, absorption in the 1.27 $\mu$m dimer band can be detected with SNR$>$1 in even some 0.1 bar cases, meaning that there could be a detectable dimer absorption signal for even a completely cloud-covered planet if the cloud deck pressure was $\geq$0.1 bars. For reflected spectra, clouds will truncate paths before they reach the surface and limit the dimer absorption for those paths. However, for partially cloudy atmospheres, dimer absorption could be detectable in the paths that do reach the surface. Additionally, because cloud albedos are typically greater than surface albedos, the presence of clouds will increase the continuum brightness levels and the brightness in the center of absorption bands. The increase in brightness has been shown to decrease the required SNR to detect and characterize O$_2$ monomer absorption in cloudy atmospheres when compared to cloud-free cases \citep{evans11}, though the effect of cloud albedo on the detectability of dimer absorption features has not been heretofore examined. Therefore, while clouds will impact the detectability of dimer features, using dimers to determine pressure and as biosignatures may still be feasible for cloudy atmospheres.

\subsection{Detectability at Different Resolving Powers} \label{sec:resolution2}

\begin{figure}
\centering
\includegraphics[width=8cm]{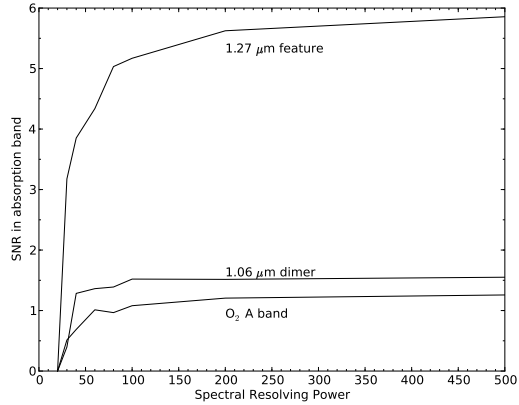}
\caption{SNRs of absorption features for the spectra shown in Figure \ref{fig:specres1}. The SNRs decrease as resolving power decreases because the absorption bands can no longer be resolved from the continuum. For R=30 and R=20 it is very difficult to identify any of the O$_2$ spectral features. \label{fig:specres2}}
\end{figure}

Figures \ref{fig:specres2} shows the SNRs for spectral features at varying resolving powers for transit transmission spectra of a 1.0 bar, 1.0x PAL O$_2$ atmosphere. The SNRs for each band are greatest at the highest resolving powers, and then gradually decrease until R$\sim$60 or 80, at which the SNRs decrease strongly. This dramatic decrease with resolving power occurs because at the lowest resolving powers, the absorption bands are indistinguishable from the continuum. Additionally, the highest flux levels in the continuum cannot be resolved at lower spectral resolving powers, decreasing the total signal. This effect can be seen most easily for the spectra with R=20, in which no absorption features can be identified.

Figures \ref{fig:specres-imaging-snrd} and \ref{fig:specres-imaging-snrp} show the SNRs for the direct imaging reflected spectra. In contrast to Figure \ref{fig:specres2}, these two figures show the required SNR to detect and characterize an absorption band, not the SNR that could be obtained with JWST. The required SNR to detect spectral features increases as resolving power decreases. However, this effect would be mitigated because the expected noise at each wavelength should decrease as resolving power decreases. The SNR required to quantify each absorption band decreases as resolving power decreases because the lowest radiance level increases. At R$<$40, however, spectral features are very difficult to identify, making these resolving powers unsuitable for detecting and characterizing O$_2$-related absorption features. 

\subsection{O$_2$ Dimer Biosignatures}

In addition to their utility as pressure probes, the 1.06 $\mu$m dimer feature and 1.27 $\mu$m feature could potentially be detectable biosignatures for nearby Earth-like planets. The O$_2$ A band has long been considered the most viable O$_2$ biosignature, but it is unlikely to be the most detectable biosignature for an Earth-like planet in transit transmission. As initially described in \citep{palle09}, lunar eclipse observations show that the 1.06 $\mu$m and 1.27 $\mu$m dimer features are more detectable than O$_2$ monomer features like the A band, which is corroborated by our model spectra and detectability calculations. The 1.27 $\mu$m O$_2$ feature has been examined as a potential biosignature for ground-based telescopes by \citet{kawahara12}, but to our knowledge detectability studies of neither the 1.06 $\mu$m dimer feature nor the 1.27 $\mu$m feature have been undertaken for JWST. Our results show that the 1.27 $\mu$m feature would be detectable with a SNR of 5 for a cloud-free Earth-analog at 5 pc.  Therefore, we conclude that O$_2$ features, especially the 1.06 $\mu$m dimer feature and the 1.27 $\mu$m feature, could be detectable biosignatures for oxygenic photosynthesis with JWST.

\subsection{Challenges to Observations}

Clouds and aerosols will make estimating pressure using dimer features more difficult. A direct imaging observation of a partially cloud-covered planet will be able to probe to the surface for a fraction of the paths, such that the dimer feature will be weaker than for a cloud-free planet. Cloud and aerosol extinction can also be wavelength dependent, which may complicate using equivalent widths or ppm flux differences to determine pressure. Nevertheless dimer features can still provide a lower bound for pressure if clouds and aerosols cannot be explicitly included in the retrieval method.

Higher H$_2$O or CO$_2$ abundances in an atmosphere could also make this method more challenging. Higher H$_2$O abundances will make it more difficult to define a continuum for the 1.06 $\mu$m dimer feature, as shown in Figures \ref{fig:h2o-test-2} and \ref{fig:h2o-test-3}. However, as discussed in Section \ref{sec:sensitivity}, the magnitude of this error should typically be less than 20\% for most cases in which the 1.06 $\mu$m dimer feature could be detectable. CO$_2$ has an absorption feature near 1.06 $\mu$m \citep{seg07}, which could make using the 1.06 $\mu$m dimer feature difficult. This could be overcome by modeling out absorption from H$_2$O and CO$_2$ or by using other dimer features to supplement information from the 1.06 $\mu$m dimer feature, such as the 1.27 $\mu$m dimer feature.

Lastly, not knowing the mixing ratio of O$_2$ will make estimating pressure difficult when only a reflected spectrum is available. This is similar to the problem in using the absorption widths of rotation-vibration features to constrain pressure. However, the 1.06 $\mu$m dimer feature is more sensitive to pressure than a monomer feature, and therefore can provide a better estimate of pressure than a monomer feature alone. Furthermore, monomer features cannot be used as an on/off pressure gauge, while dimer features can.

\section{Conclusions}

Spectrally resolved observations of O$_2$ dimer absorption can be combined with observations of the absorption by O$_2$ vibration-rotation bands to provide independent constraints on the O$_2$ concentration and the surface or cloud-top pressure in oxygenated atmospheres for planets around M dwarfs, and with low levels of CO$_2$.  Even if a precise estimate for the pressure is not possible, the presence of dimer absorption is indicative of pressures greater than $\simeq$0.5 bars in transmission spectroscopy, and greater than $\simeq$1 bars in reflected spectra.

We have shown that this method is feasible in the absence of clouds for oxygenated atmospheres.  For transmission spectroscopy this method can be used to estimate the surface or cloud-top pressure for O$_2$ concentrations $\ge$ 100\% PAL and for pressures $>$0.5 bars.  For reflected spectra this method will work if the O$_2$ partial pressure at the surface is $>$0.3 bars. Clouds will reduce the absorption strength of dimer features by limiting the paths that can probe the highest pressures in transit transmission, and by truncating paths before they reach the surface in the reflected spectra. 

JWST may be able to detect the 1.06 $\mu$m dimer feature and the 1.27 $\mu$m O$_2$ feature for an Earth analog orbiting an M dwarf in transit transmission. Thus, while not all the techniques described here will be applicable to JWST observations, a lower bound on pressure could be set for an exoplanet using an O$_2$ dimer feature. Furthermore, we showed that a direct imaging mission operating in the visible and near-infrared like TPF would require a spectral resolving power of R$>$40, and preferably higher. At R=100, a SNR of $>$30 would be required to not only detect O$_2$ related absorption features, but to also provide an estimate of an exoplanet's atmospheric pressure. 

\section*{Acknowledgments}
We thank Drake Deming for helpful dscussions on the detectability calculations, and the two anonymous reviewers for their thorough and helpful reviews that greatly improved the paper.

This work was performed by the NASA Astrobiology Institute's Virtual Planetary Laboratory, supported by the National Aeronautics and Space Administration through the NASA Astrobiology Institute under Cooperative Agreement solicitation NNH05ZDA001C.  This work has also been supported by a generous fellowship from the ARCS Seattle chapter and funding from the Astrobiology program at the University of Washington under an NSF IGERT award.

Some of the work described here was conducted at the Jet Propulsion Laboratory, California Institute of Technology, under contract with NASA.

This research has made use of NASA's Astrophysics Data System.

\section*{Author Disclosure Statement}

No competing financial interests exist.

\clearpage

\begin{table}
\begin{tabular}{|l|l|l|l|l|l|l|l|}
\hline
Pressure & fO$_2$ & \multicolumn{3}{|c|}{Transit Transmission} & \multicolumn{3}{|c|}{Secondary Eclipse} \\ \hline
Bars & PAL & O$_2$ A band & O$_2$-O$_2$ 1.06 $\mu$m & O$_2$ 1.27 $\mu$m  & O$_2$ A band & O$_2$-O$_2$ 1.06 $\mu$m & O$_2$ 1.27 $\mu$m \\ \hline
0.1 & 0.1 & 0.2 & 0.0 & 0.0 & 5.2e-04 & 0.0e+00 & 2.7e-03\\
0.1 & 0.5 & 0.5 & 0.0 & 0.9 & 1.2e-03 & 0.0e+00 & 3.4e-03\\
0.1 & 1.0 & 0.6 & 0.0 & 1.2 & 1.7e-03 & 0.0e+00 & 4.2e-03\\
0.1 & 2.0 & 0.8 & 0.0 & 1.5 & 2.4e-03 & 0.0e+00 & 5.5e-03\\
0.5 & 0.1 & 0.5 & 0.0 & 1.1 & 2.3e-03 & 0.0e+00 & 5.5e-03\\
0.5 & 0.5 & 0.8 & 0.0 & 2.5 & 5.0e-03 & 0.0e+00 & 1.2e-02\\
0.5 & 1.0 & 1.1 & 1.2 & 3.7 & 6.7e-03 & 0.0e+00 & 1.8e-02\\
0.5 & 2.0 & 1.1 & 2.3 & 5.1 & 8.5e-03 & 1.2e-02 & 2.7e-02\\
1.0 & 0.1 & 0.5 & 0.0 & 1.6 & 4.1e-03 & 0.0e+00 & 9.5e-03\\
1.0 & 0.5 & 0.8 & 0.5 & 3.9 & 7.8e-03 & 0.0e+00 & 2.6e-02\\
1.0 & 1.0 & 1.1 & 1.5 & 5.2 & 9.4e-03 & 1.3e-02 & 3.9e-02\\
1.0 & 2.0 & 1.1 & 3.4 & 7.5 & 1.1e-02 & 3.5e-02 & 6.3e-02\\
3.0 & 0.1 & 0.5 & 0.0 & 1.2 & 7.3e-03 & 0.0e+00 & 3.2e-02\\
3.0 & 0.5 & 0.8 & 0.6 & 3.5 & 1.0e-02 & 2.1e-02 & 8.4e-02\\
3.0 & 1.0 & 1.1 & 1.5 & 5.0 & 1.1e-02 & 6.2e-02 & 9.9e-02\\
3.0 & 2.0 & 1.1 & 3.5 & 7.2 & 1.2e-02 & 1.5e-01 & 1.2e-01\\
5.0 & 0.1 & 0.5 & 0.0 & 1.1 & 7.5e-03 & 0.0e+00 & 5.8e-02\\
5.0 & 0.5 & 0.8 & 0.6 & 3.3 & 9.6e-03 & 4.3e-02 & 1.1e-01\\
5.0 & 1.0 & 1.1 & 1.4 & 4.9 & 1.1e-02 & 1.2e-01 & 9.4e-02\\
5.0 & 2.0 & 1.1 & 3.3 & 7.2 & 1.2e-02 & 2.1e-01 & 1.1e-01\\
10.0 & 0.1 & 0.5 & 0.0 & 1.1 & 5.9e-03 & 0.0e+00 & 9.2e-02\\
10.0 & 0.5 & 0.8 & 0.7 & 3.4 & 7.8e-03 & 9.3e-02 & 6.1e-02\\
10.0 & 1.0 & 1.1 & 1.4 & 5.0 & 8.6e-03 & 1.7e-01 & 6.1e-02\\
10.0 & 2.0 & 1.1 & 3.4 & 7.4 & 9.0e-03 & 2.1e-01 & 5.0e-02\\
\hline
\end{tabular}
\caption{SNRs for the O$_2$ A band, the 1.06 $\mu$m dimer feature and the 1.27 $\mu$m feature for all the cases considered in both transit transmission and secondary eclipse. The calculations were done for an Earth analog orbiting an M5V star at a distance of 5 pc. The total integration time was assumed to be 10$^6$s, equal to the total amount of time spent in transit for this case over JWST's 5 year mission lifetime. The 1.06 $\mu$m dimer feature and the 1.27 $\mu$m feature could be detectable, allowing a lower limit for atmospheric pressure to be set. \label{tab:detect}}
\end{table}

\begin{table}
\begin{tabular}{|l|l|l|l|l|l|l|l|}
\hline
Pressure & fO$_2$ & \multicolumn{2}{|c|}{O$_2$ A band} & \multicolumn{2}{|c|}{1.06 $\mu$m dimer } & \multicolumn{2}{|c|}{1.27 $\mu$m  feature}\\ \hline
Bars & PAL & SNR$_D$ & SNR$_P$ & SNR$_D$ & SNR$_P$ & SNR$_D$ & SNR$_P$\\ \hline
0.1 & 0.1 & $>$100 & 3.1 & - & - & - & - \\ 
0.1 & 0.5 & 50.0 & 3.2 & - & - & $>$100 & 3.0 \\ 
0.1 & 1.0 & 35.0 & 5.4 & - & - & $>$100 & 5.1 \\ 
0.1 & 2.0 & 24.4 & 3.4 & - & - & 89.8 & 3.1 \\ 
0.5 & 0.1 & 25.1 & 3.4 & - & - & $>$100 & 3.1 \\ 
0.5 & 0.5 & 11.4 & 3.9 & - & - & 38.8 & 3.2 \\ 
0.5 & 1.0 & 8.6 & 7.3 & - & - & 23.8 & 5.6 \\ 
0.5 & 2.0 & 6.8 & 5.0 & 30.9 & 3.2 & 14.3 & 3.7 \\ 
1.0 & 0.1 & 13.4 & 3.8 & - & - & 62.0 & 3.1 \\ 
1.0 & 0.5 & 7.1 & 4.8 & - & - & 16.4 & 3.6 \\ 
1.0 & 1.0 & 5.8 & 9.1 & 29.0 & 5.4 & 9.9 & 6.9 \\ 
1.0 & 2.0 & 5.0 & 6.1 & 9.6 & 3.7 & 5.9 & 5.4 \\ 
3.0 & 0.1 & 6.2 & 5.3 & - & - & 12.8 & 3.8 \\ 
3.0 & 0.5 & 4.4 & 6.9 & 16.5 & 3.4 & 4.3 & 7.2 \\ 
3.0 & 1.0 & 3.9 & 13.2 & 4.9 & 8.1 & 2.9 & 26.7 \\ 
3.0 & 2.0 & 3.5 & 9.4 & 1.9 & 18.9 & 1.7 & $>$100 \\ 
5.0 & 0.1 & 5.0 & 6.3 & - & - & 6.5 & 5.0 \\ 
5.0 & 0.5 & 3.8 & 8.6 & 6.8 & 4.1 & 2.7 & 20.9 \\ 
5.0 & 1.0 & 3.4 & 17.7 & 2.3 & 18.1 & 1.9 & $>$100 \\ 
5.0 & 2.0 & 3.0 & 15.2 & 1.3 & $>$100 & 1.3 & $>$100 \\ 
10.0 & 0.1 & 4.1 & 8.1 & - & - & 3.0 & 13.7 \\ 
10.0 & 0.5 & 3.0 & 15.2 & 2.3 & 10.4 & 1.8 & $>$100 \\ 
10.0 & 1.0 & 2.7 & 44.9 & 1.4 & $>$100 & 1.2 & $>$100 \\ 
10.0 & 2.0 & 2.4 & 61.6 & 1.0 & $>$100 & 1.0 & $>$100 \\ 
\hline
\end{tabular}
\caption{SNRs for detection (SNR$_D$) and to obtain precision of 3$\sigma$ in the center of the band (SNR$_P$) for the O$_2$ A band, the 1.06 $\mu$m dimer feature and the 1.27 $\mu$m feature at pressures ranging from 0.1 to 10.0 bars and O$_2$ concentrations ranging from 0.1 to 2.0 times PAL O$_2$ for direct imaging reflected spectra.  \label{tab:TPF}}
\end{table}

\clearpage

\clearpage

\begin{thebibliography}

\bibitem[Acarreta et al.(2004)]{acaretta04} Acarreta, J.~R., de 
Haan, J.~F., Stammes, P.\ 2004.\ Cloud pressure retrieval using the 
O$_{2}$-O$_{2}$ absorption band at 477 nm.\ Journal of Geophysical Research 
(Atmospheres) 109, 5204. 

\bibitem[Auer and Standish(2000)]{auer00} Auer, L.~H., 
Standish, E.~M.\ 2000.\ Astronomical Refraction: Computational Method for 
All Zenith Angles.\ The Astronomical Journal 119, 2472-2474. 

\bibitem[Baldridge et al.(2009)]{baldridge2009} Baldridge, A. M., S.J. Hook, C.I. Grove and G. Rivera, 2009.. The ASTER Spectral Library Version 2.0. Remote Sensing of Environment, vol 113, pp. 711-715. 

\bibitem[Barton and Scott(1986)]{barton86} Barton, I.~J., Scott, 
J.~C.\ 1986.\ Remote measurement of surface pressure using absorption in 
the oxygen A-band.\ Applied Optics 25, 3502-3507. 


\bibitem[Bean et al.(2010)]{bean10} Bean, J.~L., Miller-Ricci 
Kempton, E., Homeier, D.\ 2010.\ A ground-based transmission spectrum of 
the super-Earth exoplanet GJ 1214b.\ Nature 468, 669-672. 

\bibitem[Belu et 
al.(2011)]{belu11} Belu, A.~R., Selsis, F., Morales, J.-C., Ribas, I., Cossou, C., Rauer, H.\ 2011.\ Primary and secondary eclipse spectroscopy with JWST: exploring the exoplanet parameter space.\ Astronomy and Astrophysics 525, A83. 


\bibitem[Brown(2001)]{brown01} Brown, T.~M.\ 2001.\ 
Transmission Spectra as Diagnostics of Extrasolar Giant Planet 
Atmospheres.\ The Astrophysical Journal 553, 1006-1026. 

\bibitem[Carey et al.(2012)]{carey12} Carey, S., and 12 
colleagues 2012.\ Absolute photometric calibration of IRAC: lessons learned 
using nine years of flight data.\ Society of Photo-Optical Instrumentation 
Engineers (SPIE) Conference Series 8442, . 


\bibitem[Catling et al.(2010)]{catling10} Catling, D.~C., Claire, 
M.~W., Zahnle, K.~J., Quinn, R.~C., Clark, B.~C., Hecht, M.~H., Kounaves, 
S.\ 2010.\ Atmospheric origins of perchlorate on Mars and in the Atacama.\ 
Journal of Geophysical Research (Planets) 115, 0. 


\bibitem[Chamberlain et al.(2013)]{chamb13} Chamberlain, S., 
Bailey, J., Crisp, D., Meadows, V.\ 2013.\ Ground-based near-infrared 
observations of water vapour in the Venus troposphere.\ Icarus 222, 
364-378. 


\bibitem[Chamberlain et al.(2006)]{chamb06} Chamberlain, S.~A., 
Bailey, J.~A., Crisp, D.\ 2006.\ Mapping Martian Atmospheric Pressure with 
Ground-Based Near Infrared Spectroscopy.\ Publications of the Astronomical 
Society of Australia 23, 119-124. 


\bibitem[Charbonneau et al.(2002)]{char02} Charbonneau, D., 
Brown, T.~M., Noyes, R.~W., Gilliland, R.~L.\ 2002.\ Detection of an 
Extrasolar Planet Atmosphere.\ The Astrophysical Journal 568, 377-384. 

\bibitem[Clark et al.(2007)]{clark2007} Clark, R.N., Swayze, G.A., Wise, R., Livo, E., Hoefen, T., Kokaly, R., Sutley, S.J., 2007, USGS digital spectral library splib06a: U.S. Geological Survey, Digital Data Series 231.

\bibitem[Crisp(1997)]{crisp97} Crisp, D.\ 1997.\ Absorption of 
sunlight by water vapor in cloudy conditions: A partial explanation for the 
cloud absorption anomaly.\ Geophysical Research Letters 24, 571-574. 

\bibitem[Crisp et al.(2012)]{crisp12} Crisp, D., and 33 
colleagues 2012.\ The ACOS CO$_{2}$ retrieval algorithm - Part II: Global 
X$_{CO_2}$ data characterization.\ Atmospheric Measurement Techniques 5, 
687-707. 


\bibitem[Crow et al.(2011)]{crow11} Crow, C.~A., McFadden, 
L.~A., Robinson, T., Meadows, V.~S., Livengood, T.~A., Hewagama, T., Barry, 
R.~K., Deming, L.~D., Lisse, C.~M., Wellnitz, D.\ 2011.\ Views from EPOXI: 
Colors in Our Solar System as an Analog for Extrasolar Planets.\ The 
Astrophysical Journal 729, 130. 

\bibitem[Deming et al.(2009)]{deming09} Deming, D., and 11 
colleagues 2009.\ Discovery and Characterization of Transiting Super Earths 
Using an All-Sky Transit Survey and Follow-up by the James Webb Space 
Telescope.\ Publications of the Astronomical Society of the Pacific 121, 
952-967. 

\bibitem[Deming et al.(2013)]{deming13} Deming, D., and 20 
colleagues 2013.\ Infrared Transmission Spectroscopy of the Exoplanets HD 
209458b and XO-1b Using the Wide Field Camera-3 on the Hubble Space 
Telescope.\ The Astrophysical Journal 774, 95. 

\bibitem[Domagal-Goldman et al.(2011)]{domagal11} 
Domagal-Goldman, S.~D., Meadows, V.~S., Claire, M.~W., Kasting, J.~F.\ 
2011.\ Using Biogenic Sulfur Gases as Remotely Detectable Biosignatures on 
Anoxic Planets.\ Astrobiology 11, 419-441. 

\bibitem[Evans et al.(2011)]{evans11} Evans, N., Meadows, 
V.~S., Domagal-Goldman, S.~D.\ 2011.\ Exploring The Detectability of 
Terrestrial Exoplanet Characteristics.\ Bulletin of the American 
Astronomical Society 43, \#343.09. 


\bibitem[Forget et al.(2007)]{forget07} Forget, F., Spiga, A., 
Dolla, B., Vinatier, S., Melchiorri, R., Drossart, P., Gendrin, A., 
Bibring, J.-P., Langevin, Y., Gondet, B.\ 2007.\ Remote sensing of surface 
pressure on Mars with the Mars Express/OMEGA spectrometer: 1. Retrieval 
method.\ Journal of Geophysical Research (Planets) 112, 8. 


\bibitem[Garc{\'{\i}}a Mu{\~n}oz et al.(2012)]{garcia12} 
Garc{\'{\i}}a Mu{\~n}oz, A., Zapatero Osorio, M.~R., Barrena, R., 
Monta{\~n}{\'e}s-Rodr{\'{\i}}guez, P., Mart{\'{\i}}n, E.~L., Pall{\'e}, E.\ 
2012.\ Glancing Views of the Earth: From a Lunar Eclipse to an Exoplanetary 
Transit.\ The Astrophysical Journal 755, 103. 


\bibitem[Gardner et al.(2006)]{gardner06} Gardner, J.~P., and 22 
colleagues 2006.\ The James Webb Space Telescope.\ Space Science Reviews 
123, 485-606. 


\bibitem[Gray(1966)]{gray66} Gray, L.~D.\ 1966.\ Transmission 
of the Atmosphere of Mars in the Region of 2 $\mu$m.\ Icarus 5, 390. 


\bibitem[Greenblatt et al.(1990)]{greenblatt90} Greenblatt, G.~D., 
Orlando, J.~J., Burkholder, J.~B., Ravishankara, A.~R.\ 1990.\ Absorption 
measurements of oxygen between 330 and 1140 nm.\ Journal of Geophysical 
Research 95, 18577-18582. 

\bibitem[Hauschildt et al.(1999)]{hauschildt99} Hauschildt, P.~H., 
Allard, F., Baron, E.\ 1999.\ The NextGen Model Atmosphere Grid for 
3000$<$=T\_eff$<$=10,000 K.\ The Astrophysical Journal 512, 377-385. 

\bibitem[Hubbard et al.(2001)]{hubbard01} Hubbard, W.~B., 
Fortney, J.~J., Lunine, J.~I., Burrows, A., Sudarsky, D., Pinto, P.\ 2001.\ 
Theory of Extrasolar Giant Planet Transits.\ The Astrophysical Journal 560, 
413-419. 


\bibitem[Ignatiev et al.(2009)]{ign09} Ignatiev, N.~I., 
Titov, D.~V., Piccioni, G., Drossart, P., Markiewicz, W.~J., Cottini, V., 
Roatsch, T., Almeida, M., Manoel, N.\ 2009.\ Altimetry of the Venus cloud 
tops from the Venus Express observations.\ Journal of Geophysical Research 
(Planets) 114, 0. 

\bibitem[Kaltenegger and Traub(2009)]{kalt09} Kaltenegger, L., 
Traub, W.~A.\ 2009.\ Transits of Earth-like Planets.\ The Astrophysical 
Journal 698, 519-527. 

\bibitem[Kaplan et al.(1964)]{kaplan64} Kaplan, L.~D., 
M{\"u}nch, G., Spinrad, H.\ 1964.\ An Analysis of the Spectrum of Mars..\ 
The Astrophysical Journal 139, 1. 

\bibitem[Kasting et al.(2010)]{traub10}  Kasting, J. F., W. A. Traub, et al., Exoplanet characterization and the search for life, White paper for Astronomy and Astrophysics Decadal Survey, available at: 
\url{http://sites.nationalacademies.org/BPA/BPA_050603#planetarysystems}

\bibitem[Kawahara et al.(2012)]{kawahara12} Kawahara, H., Matsuo, 
T., Takami, M., Fujii, Y., Kotani, T., Murakami, N., Tamura, M., Guyon, O.\ 
2012.\ Can Ground-based Telescopes Detect the Oxygen 1.27 {$\mu$}m 
Absorption Feature as a Biomarker in Exoplanets?.\ The Astrophysical 
Journal 758, 13. 

\bibitem[K{\"o}hler et al.(2005)]{kohler05} K{\"o}hler, J., 
Melf, M., Posselt, W., Holota, W., te Plate, M.\ 2005.\ Optical design of 
the near-infrared spectrograph NIRSpec.\ Society of Photo-Optical 
Instrumentation Engineers (SPIE) Conference Series 5962, 563-574. 

\bibitem[Kump(2008)]{kump08} Kump, L.~R.\ 2008.\ The rise of 
atmospheric oxygen.\ Nature 451, 277-278. 


\bibitem[Lafreniere et al.(2013)]{laf13} Lafreniere, D., 
Doyon, R., FGS/NIRISS, NIRCam, MIRI, NIRSpec Science Teams 2013.\ The 
Science Potential of JWST for Exoplanet Studies.\ American Astronomical 
Society Meeting Abstracts 221, \#135.06. 


\bibitem[Mat{\'e} et al.(1999)]{mate99} Mat{\'e}, B., Lugez, 
C., Fraser, G.~T., Lafferty, W.~J.\ 1999.\ Absolute intensities for the 
O$_{2}$ 1.27 {$\mu$}m continuum absorption.\ Journal of Geophysical 
Research 104, 30585-30590. 


\bibitem[Meadows and Crisp(1996)]{mead96} Meadows, V.~S., 
Crisp, D.\ 1996.\ Ground-based near-infrared observations of the Venus 
nightside: The thermal structure and water abundance near the surface.\ 
Journal of Geophysical Research 101, 4595-4622. 

\bibitem[Mitchell and O'Brien(1987)]{mitchell87} Mitchell, R.~M., 
O'Brien, D.~M.\ 1987.\ Error Estimates for Passive Satellite Measurement of 
Surface Pressure Using Absorption in the A Band of Oxygen..\ Journal of 
Atmospheric Sciences 44, 1981-1990. 

\bibitem[Pall{\'e} et al.(2009)]{palle09} Pall{\'e}, E., 
Zapatero Osorio, M.~R., Barrena, R., Monta{\~n}{\'e}s-Rodr{\'{\i}}guez, P., 
Mart{\'{\i}}n, E.~L.\ 2009.\ Earth's transmission spectrum from lunar 
eclipse observations.\ Nature 459, 814-816. 

\bibitem[Pierrehumbert(2010)]{pierrehumbert10} Pierrehumbert, R.~T.\ 
2010.\ Principles of Planetary Climate.\ Principles of Planetary Climate, 
by R.~T Pierrehumbert.~ Cambridge, UK: Cambridge University Press.~ISBN: 
9780521865562, 2010 . 

\bibitem[Pont et al.(2008)]{pont08} Pont, F., Knutson, H., 
Gilliland, R.~L., Moutou, C., Charbonneau, D.\ 2008.\ Detection of 
atmospheric haze on an extrasolar planet: the 0.55-1.05 {$\mu$}m 
transmission spectrum of HD 189733b with the Hubble Space Telescope.\ Monthly 
Notices of the Royal Astronomical Society 385, 109-118. 

\bibitem[Rauer et 
al.(2011)]{rauer11} Rauer, H., Gebauer, S., Paris, P.~V., Cabrera, J., Godolt, M., Grenfell, J.~L., Belu, A., Selsis, F., Hedelt, P., Schreier, F.\ 2011.\ Potential biosignatures in super-Earth atmospheres. I. Spectral appearance of super-Earths around M dwarfs.\ Astronomy and Astrophysics 529, A8. 

\bibitem[Rothman et al.(2009)]{rothman09} Rothman, L.~S., and 42 
colleagues 2009.\ The HITRAN 2008 molecular spectroscopic database.\ 
Journal of Quantitative Spectroscopy and Radiative Transfer 110, 533-572. 

\bibitem[Seager and Sasselov(2000)]{seager00} Seager, S., 
Sasselov, D.~D.\ 2000.\ Theoretical Transmission Spectra during Extrasolar 
Giant Planet Transits.\ The Astrophysical Journal 537, 916-921. 

\bibitem[Segura et al.(2005)]{segura05} Segura, A., Kasting, 
J.~F., Meadows, V., Cohen, M., Scalo, J., Crisp, D., Butler, R.~A.~H., 
Tinetti, G.\ 2005.\ Biosignatures from Earth-Like Planets Around M Dwarfs.\ 
Astrobiology 5, 706-725. 

\bibitem[Segura et 
al.(2007)]{seg07} Segura, A., Meadows, V.~S., Kasting, J.~F., Crisp, D., Cohen, M.\ 2007.\ Abiotic formation of O$_{2}$ and O$_{3}$ in high-CO$_{2}$ terrestrial atmospheres.\ Astronomy and Astrophysics 472, 665-679. 

\bibitem[Slanina et al.(1994)]{slanina94}Slanina, Z., Uhlík, F., De Almeida, W., Hinchliffe, A.\ 1994.\ A computational thermodynamic evaluation of the altitude profiles of (N$_2$)$_2$, N$_2$-O$_2$ and (O$_2$)$_2$ in the Earth's atmosphere.\ Thermochimica Acta 231, 55-60.


\bibitem[Spiga et al.(2007)]{spiga07} Spiga, A., Forget, F., 
Dolla, B., Vinatier, S., Melchiorri, R., Drossart, P., Gendrin, A., 
Bibring, J.-P., Langevin, Y., Gondet, B.\ 2007.\ Remote sensing of surface 
pressure on Mars with the Mars Express/OMEGA spectrometer: 2. 
Meteorological maps.\ Journal of Geophysical Research (Planets) 112, 8. 

\bibitem[Uhlik et al.(1993)]{uhlik93} Uhlik, F, Slanina, Z, and Hinchliffe, A.\ 1993.\ Gas-phase association of O2: a computation thermodynamic study .\ Thermochimica Acta, 228.

\bibitem[Vidal-Madjar et al.(2003)]{vidal03} Vidal-Madjar, A., 
Lecavelier des Etangs, A., D{\'e}sert, J.-M., Ballester, G.~E., Ferlet, R., 
H{\'e}brard, G., Mayor, M.\ 2003.\ An extended upper atmosphere around the 
extrasolar planet HD209458b.\ Nature 422, 143-146. 

\bibitem[Wakeford et al.(2013)]{wakeford13} Wakeford, H.~R., and 
15 colleagues 2013.\ HST hot Jupiter transmission spectral survey: 
detection of water in HAT-P-1b from WFC3 near-IR spatial scan 
observations.\ Monthly Notices of the Royal Astronomical Society 435, 
3481-3493. 

\bibitem[Zerkle et al.(2012)]{zerkle12} Zerkle, A.~L., Claire, 
M.~W., Domagal-Goldman, S.~D., Farquhar, J., Poulton, S.~W.\ 2012.\ A 
bistable organic-rich atmosphere on the Neoarchaean Earth.\ Nature 
Geoscience 5, 359-363. 

\end{thebibliography}
\end{document}